\documentclass[12pt,twoside,dvips]{article}
\usepackage{amssymb}
\usepackage{amsfonts}
\usepackage{amsmath}
\usepackage[spanish,english]{babel}
\usepackage{graphicx}

\begin{document}

\title{$Z$ and $Z^{\prime }$ decays with and without FCNC in $331$ models,}
\author{A. Carcamo, R. Mart\'{\i}nez$\thanks{%
e-mail: remartinezm@unal.edu.co}$ \ and F. Ochoa$\thanks{%
e-mail: faochoap@unal.edu.co}$ \and Departamento de F\'{\i}sica, Universidad
Nacional, \\
Bogot\'{a}-Colombia}
\maketitle

\begin{abstract}
In the context of the 331 models, we consider constraints on the extra
neutral boson $Z^{\prime}$ predicted by the model, where three different
quark family assignments are identified. Using the ansatz of Matsuda as an
specific texture for the quark mass matrices, we obtain allowed regions
associated with the Z-Z' mixing angle, the mass of the $Z^{\prime}$ boson
and the parameter $\beta$ which determines different 331 models. The $Z_{1}$
and $Z_{2}$ decays with and without flavor changing are also considered. The
flavor changing decays of the $Z_{1}$ boson into quarks at tree level are
highly suppressed by the $Z-Z^{\prime}$ mixing angle, obtaining the same
order of magnitude as the standard model prediction at one loop level. The $%
Z_{2}$ decay widths are calculated with and without flavor changing, where
oblique radiative corrections at one loop accounts for about $1\%-4\%$
deviations.
\end{abstract}

\vspace{-0.3cm}

\vspace{-5mm}

\section{Introduction}

In most of extensions of the standard model (SM), new massive and neutral
gauge bosons, called $Z^{\prime }$, are predicted. The phenomenological
features that arise about such boson have been subject of extensive studies in
the literature \cite{zprima}, whose presence is sensitive to experimental
observations at low and high energies, and will be of great interest in the
next generation of colliders (LHC, ILC) \cite{godfrey}. In particular, it is
possible to study some phenomenological features associated with this extra
neutral gauge boson through models with gauge symmetry $SU(3)_{c}\otimes
SU(3)_{L}\otimes U(1)_{X},$ also called 331 models. These models arise as an
interesting alternative to explain the origin of generations \cite{Frampton2}%
, where the three families are required in order to cancel chiral anomalies
completely \cite{anomalias}. An additional motivation to study these kinds
of models comes from the fact that they can also predict the charge
quantization for a three family model even when neutrino masses are added 
\cite{Pires}.

Although cancellation of anomalies leads to some required conditions \cite%
{fourteen}, such criterion alone still permits an infinite number of 331
models. In these models, the electric charge is defined in general as a
linear combination of the diagonal generators of the group

\begin{equation}
Q=T_{3}+\beta T_{8}+XI,  \label{charge}
\end{equation}%
where the value of the $\beta $ parameter determines the fermion assignment
and, more specifically, the electric charges of the exotic spectrum. Hence,
it is customary to use this quantum number to classify the different 331
models. If we want to avoid exotic charges we are led to only two different
models i.e. $\beta =\pm 1/\sqrt{3}$ \cite{fourteen, twelve}. An extensive
and detailed study of models with $\beta $ arbitrary has been carried out
in ref. \cite{331us} for the scalar sector and in ref. \cite{beta-arbitrary}
for the fermionic and gauge sector.

The group structure of these models leads, along with the SM-neutral boson $%
Z,$ to the prediction of an additional current associated with a new
neutral boson $Z^{\prime }.$ It is possible to study the low energy
deviations of the $Z-$pole observables through a precision fit of the $Z-Z^{\prime }$ mixing \cite{ten,Frampton}, which may
provides indirect constraints on the free parameters of the model (including 
$\beta $). Unlike the $Z$-boson whose couplings are family independent and
the weak interactions at low energy are of universal character, the
couplings of $Z^{\prime }$ are different for the three families due to the $%
U(1)_{X}$ values to each of them. In the quark sector each of the
331 families in the weak basis can be assigned in three different ways into
mass eigenstates. In this way, in a phenomenological analysis, the allowed
region associated with the $Z-Z^{\prime }$ mixing angle and the physical
mass $M_{Z^{\prime }}$ of the extra neutral boson will depend on the family
assignment. This study was carried out in ref. \cite{family-dependence} for
the two main versions of the 331 models corresponding to $\beta =-\sqrt{3}$ 
\cite{ten} and $\beta =-\frac{1}{\sqrt{3}}$ \cite{twelve}, and in ref. \cite%
{evaluation} for $\beta $ arbitrary.

In addition, the study of rare decays provides a framework to evaluate any
new physics beyond the SM (for a recent review on the rare Z decays, see
ref. \cite{Perez1}). In particular, the SM may induce FCNC in the $Z$ decay
by introducing either one loop corrections or effective couplings with
dimension 6 \cite{ghosal}. In the SM, the flavor changing processes in the Z gauge sector are forbidden at tree level even though we considered rotations of the fermions from weak to mass eigenstates. In the context of models beyond the SM with extra neutral
currents, the $Z-Z^{\prime }$ mixing produces small deviations that break
the universality feature, and induces FCNC at tree level in $Z$ decays \cite%
{chiang, sher,Perez2} when rotations between weak and mass eigenstates are
implemented. Such flavor changing couplings are model dependent \cite%
{langaker1}, and its realization could be verified in the future $e^{+}e^{-}$
lineal colliders at the TeV scale (TESLA \cite{tesla}). In the case of 331
models, FCNC at tree level due to the $Z-Z^{\prime }$ mixing arise only in
the quark sector, while the couplings of the leptons with $Z$ and $Z^{\prime
}$ are universal of families (ref. \cite{sher} considers new 331 models with
FCNC in the lepton sector). We take specific textures for the quark mass
matrices in agreement with the current data on the CKM mixing angles. The
assignments for texture on mass fermions have been broadly discussed in the
literature \cite{matsuda, texture}.

In this work we report a phenomenological study of the 331-extra neutral
boson. First, we consider indirect limits at the $Z$ resonance for models
with $\beta $ arbitrary, including linear combinations among the quark
families. We adopt the texture structure proposed in ref. \cite{matsuda} in
order to obtain allowed regions for the $Z-Z^{\prime }$ mixing angle, the
mass of the $Z^{\prime }$ boson and the values of $\beta $ for three
different assignments of the quark families \cite{Mohapatra} in mass
eigenstates. The above analysis is performed through a $\chi ^{2}$
statistics at 95\% CL including correlation data among the observables.
Later, we study high energy corrections of the $Z^{\prime }$ decay in the
framework of models with $\beta =1/\sqrt{3}$. These analysis are performed
considering oblique corrections at one loop level, where running coupling
constants at $M_{Z^{\prime }}$ scale are taken into account. We also
calculate the FCNC contributions in the $Z$ and $Z^{\prime }$ decays.

\section{The 331 spectrum for $\protect\beta$ arbitrary\label%
{sec:331-spectrum}}

The 331 fermionic structures for three families is shown in table \ref%
{tab:espectro} for $\beta $ arbitrary, where all leptons transform as $(%
\mathbf{1,3,X}_{\ell }^{L})$ and $(\mathbf{1,1,X}_{\ell ^{\prime }}^{R})$
under $\left( SU(3)_{c},SU(3)_{L},U(1)_{X}\right) ,$ with $\mathbf{X}_{\ell
}^{L}$ and $\mathbf{X}_{\ell ^{\prime }}^{R}$ the $U(1)_{X}$ values
associated with the left- and right-handed leptons, respectively; while the
quarks transform as $(\mathbf{3,3}^{\ast }\mathbf{,X}_{q_{m^{\ast }}}^{L})$, 
$(\mathbf{3}^{\ast }\mathbf{,1,X}_{q_{m^{\ast }}^{\prime }}^{R})$ for the
first two families, and $(\mathbf{3,3,X}_{q_{3}}^{L})$, $(\mathbf{3}^{\ast }%
\mathbf{,1,X}_{q_{3}^{\prime }}^{R})$ for the third family, where $\mathbf{X}%
_{q_{m^{\ast }}}^{L},\mathbf{X}_{q_{3}}^{L}$ and $\mathbf{X}_{q_{m^{\ast
}}^{\prime }}^{R},\mathbf{X}_{q_{3}^{\prime }}^{R}$ correspond to the $%
U(1)_{X}$ values for left- and right-handed quarks. We denote $\mathbf{X}%
_{q_{3}}^{L}$ and $\mathbf{X}_{q_{m^{\ast }}}^{L}$ as the values associated
with the $SU(3)_{L}$ space under $\mathbf{3}$ and $\mathbf{3}
^{\ast }$ representations, respectively. The quantum numbers $\mathbf{X}_{\psi }$ for each
representation are given in the third column from table \ref{tab:espectro},
where the definition of the electric charge in Eq. \ref{charge} has been
used, demanding charges of $2/3$ and $-1/3$ to the up- and down-type quarks,
respectively, and charges of -1,0 for the charged and neutral leptons. We
recognize three different possibilities to assign the physical quarks in
each family representation as shown in table \ref{tab:combination}. At low
energy, the three models from table \ref{tab:combination} are equivalent and
there is not any phenomenological feature that allows us to detect
differences between them. In fact, they must reduce to the SM which is an
universal family model in $SU(2)_{L}.$ However, through the couplings of the
three families to the additional neutral current ($Z^{\prime }$) and the
introduction of a mixing angle between $Z$ and $Z^{\prime },$ it is possible
to recognize differences among the three models at the electroweak scale.

\begin{table}[tbp]
\begin{center}
\begin{equation*}
\begin{tabular}{c|c|c}
\hline
$representation$ & $Q_{\psi }$ & $X_{\psi }$ \\ \hline\hline
$\ 
\begin{tabular}{c}
$q_{m^{\ast }L}=\left( 
\begin{array}{c}
d_{m^{\ast }} \\ 
-u_{m^{\ast }} \\ 
J_{m^{\ast }}%
\end{array}%
\right) _{L}\mathbf{3}^{\ast }$ \\ 
\\ 
\\ 
$d_{m^{\ast }R};$ $u_{m^{\ast }R};$ $J_{m^{\ast }R}:\mathbf{1}$%
\end{tabular}%
\ $ & 
\begin{tabular}{c}
$\left( 
\begin{array}{c}
-\frac{1}{3} \\ 
\frac{2}{3} \\ 
\frac{1}{6}+\frac{\sqrt{3}\beta }{2}%
\end{array}%
\right) $ \\ 
\\ 
$-\frac{1}{3};$ $\frac{2}{3};$ $\frac{1}{6}+\frac{\sqrt{3}}{2}\beta $%
\end{tabular}
& 
\begin{tabular}{c}
\\ 
$X_{q_{m^{\ast }}}^{L}=\frac{1}{6}+\frac{\beta }{2\sqrt{3}}$ \\ 
\\ 
\\ 
$X_{d_{m^{\ast }},u_{m^{\ast }},J_{m^{\ast }}}^{R}=-\frac{1}{3},\frac{2}{3},%
\frac{1}{6}+\frac{\sqrt{3}}{2}\beta $%
\end{tabular}
\\ \hline\hline
\begin{tabular}{c}
$q_{3L}=\left( 
\begin{array}{c}
u_{3} \\ 
d_{3} \\ 
J_{3}%
\end{array}%
\right) _{L}:\mathbf{3}$ \\ 
\\ 
$u_{3R};$ $d_{3R};$ $J_{3R}:\mathbf{1}$%
\end{tabular}
& 
\begin{tabular}{c}
$\left( 
\begin{array}{c}
\frac{2}{3} \\ 
-\frac{1}{3} \\ 
\frac{1}{6}-\frac{\sqrt{3}\beta }{2}%
\end{array}%
\right) $ \\ 
\\ 
$\frac{2}{3};$ $-\frac{1}{3};$ $\frac{1}{6}-\frac{\sqrt{3}\beta }{2}$%
\end{tabular}
& 
\begin{tabular}{c}
\\ 
$X_{q^{(3)}}^{L}=\frac{1}{6}-\frac{\beta }{2\sqrt{3}}$ \\ 
\\ 
\\ 
$X_{u_{3},d_{3},J_{3}}^{R}=\frac{2}{3},-\frac{1}{3},\frac{1}{6}-\frac{\sqrt{3%
}\beta }{2}$%
\end{tabular}
\\ \hline\hline
\begin{tabular}{c}
$\ell _{jL}=\left( 
\begin{array}{c}
\nu _{j} \\ 
e_{j} \\ 
E_{j}^{-Q_{1}}%
\end{array}%
\right) _{L}:\mathbf{3}$ \\ 
\\ 
$e_{jR};$ $E_{jR}^{-Q_{1}}$%
\end{tabular}
& 
\begin{tabular}{c}
$\left( 
\begin{array}{c}
0 \\ 
-1 \\ 
-\frac{1}{2}-\frac{\sqrt{3}\beta }{2}%
\end{array}%
\right) $ \\ 
\\ 
$-1;$ $-\frac{1}{2}-\frac{\sqrt{3}\beta }{2}$%
\end{tabular}
& 
\begin{tabular}{c}
\\ 
$X_{\ell _{j}}^{L}=-\frac{1}{2}-\frac{\beta }{2\sqrt{3}}$ \\ 
\\ 
\\ 
$X_{e_{j},E_{j}}^{R}=-1,$ $-\frac{1}{2}-\frac{\sqrt{3}\beta }{2}$%
\end{tabular}
\\ \hline
\end{tabular}%
\end{equation*}%
\end{center}
\caption{\textit{Fermionic content for three generations with\ }$\protect%
\beta \ $\textit{arbitrary. We take} $m^{\ast }=1,2$, \textit{and} $j=1,2,3$}
\label{tab:espectro}
\end{table}

For the scalar sector described by Table \ref{tab:quince}, we introduce the
triplet field $\chi $ with vacuum expectation value (VEV) $\left\langle \chi
\right\rangle ^{T}=\left( 0,0,\nu _{\chi }\right) $, which induces
masses to the third fermionic components. In the second transition it is
necessary to introduce two triplets$\;\rho $ and $\eta $ with VEV $%
\left\langle \rho \right\rangle ^{T}=\left( 0,\nu _{\rho },0\right) $ and $%
\left\langle \eta \right\rangle ^{T}=\left( \nu _{\eta },0,0\right) $ in
order to give masses to the quarks of type up and down, respectively.

\begin{table}[tbp]
\begin{equation*}
\begin{tabular}{c|c|c}
\hline
Representation $A$ & Representation $B$ & Representation $C$ \\ \hline\hline
$%
\begin{tabular}{c}
$q_{mL}=\left( 
\begin{array}{c}
d,s \\ 
-u,-c \\ 
J_{1},J_{2}%
\end{array}%
\right) _{L}:\mathbf{3}^{\ast }$ \\ 
$q_{3L}=\left( 
\begin{array}{c}
t \\ 
b \\ 
J_{3}%
\end{array}%
\right) _{L}:\mathbf{3}$%
\end{tabular}%
\ $ & $%
\begin{tabular}{c}
$q_{mL}=\left( 
\begin{array}{c}
d,b \\ 
-u,-t \\ 
J_{1},J_{3}%
\end{array}%
\right) _{L}:\mathbf{3}^{\ast }$ \\ 
$q_{3L}=\left( 
\begin{array}{c}
c \\ 
s \\ 
J_{2}%
\end{array}%
\right) _{L}:\mathbf{3}$%
\end{tabular}%
\ $ & $%
\begin{tabular}{c}
$q_{mL}=\left( 
\begin{array}{c}
s,b \\ 
-c,-t \\ 
J_{2},J_{3}%
\end{array}%
\right) _{L}:\mathbf{3}^{\ast }$ \\ 
$q_{3L}=\left( 
\begin{array}{c}
u \\ 
d \\ 
J_{1}%
\end{array}%
\right) _{L}:\mathbf{3}$%
\end{tabular}%
\ $ \\ \hline
\end{tabular}%
\end{equation*}%
\caption{\textit{Three different assignments for the $SU(3)_{L}$ family
representation of quarks}}
\label{tab:combination}
\end{table}

\begin{table}[tbp]
\begin{center}
$%
\begin{tabular}{c|c|c}
\hline
& $Q_{\Phi }$ & $X_{\Phi }$ \\ \hline\hline
$\chi =\left( 
\begin{array}{c}
\chi _{1}^{\pm Q_{1}} \\ 
\chi _{2}^{\pm Q_{2}} \\ 
\xi _{\chi }+\nu _{\chi }\pm i\zeta _{\chi }%
\end{array}%
\right) $ & $\left( 
\begin{array}{c}
\pm \left( \frac{1}{2}+\frac{\sqrt{3}\beta }{2}\right) \\ 
\pm \left( -\frac{1}{2}+\frac{\sqrt{3}\beta }{2}\right) \\ 
0%
\end{array}%
\right) $ & $\frac{\beta }{\sqrt{3}}$ \\ \hline
$\rho =\left( 
\begin{array}{c}
\rho _{1}^{\pm } \\ 
\xi _{\rho }+\nu _{\rho }\pm i\zeta _{\rho } \\ 
\rho _{3}^{\mp Q_{2}}%
\end{array}%
\right) $ & $\left( 
\begin{array}{c}
\pm 1 \\ 
0 \\ 
\mp \left( -\frac{1}{2}+\frac{\sqrt{3}\beta }{2}\right)%
\end{array}%
\right) $ & $\frac{1}{2}-\frac{\beta }{2\sqrt{3}}$ \\ \hline
$\eta =\left( 
\begin{array}{c}
\xi _{\eta }+\nu _{\eta }\pm i\zeta _{\eta } \\ 
\eta _{2}^{\mp } \\ 
\eta _{3}^{\mp Q_{1}}%
\end{array}%
\right) $ & $\left( 
\begin{array}{c}
0 \\ 
\mp 1 \\ 
\mp \left( \frac{1}{2}+\frac{\sqrt{3}\beta }{2}\right)%
\end{array}%
\right) $ & $-\frac{1}{2}-\frac{\beta }{2\sqrt{3}}$ \\ \hline
\end{tabular}%
\ \ $%
\end{center}
\caption{\textit{Scalar spectrum that break the 331 symmetry and give masses
to the fermions.}}
\label{tab:quince}
\end{table}

In the gauge boson spectrum associated with the  $SU(3)_{L}\otimes
U(1)_{X}$ gauge group, we are just interested in the physical neutral sector that
corresponds to the photon, $Z$ and $Z^{\prime },$ which are written in terms
of the electroweak basis for $\beta $ arbitrary as \cite{beta-arbitrary}

\begin{eqnarray}
A_{\mu } &=&S_{W}W_{\mu }^{3}+C_{W}\left( \beta T_{W}W_{\mu }^{8}+\sqrt{%
1-\beta ^{2}T_{W}^{2}}B_{\mu }\right) ,  \notag \\
Z_{\mu } &=&C_{W}W_{\mu }^{3}-S_{W}\left( \beta T_{W}W_{\mu }^{8}+\sqrt{%
1-\beta ^{2}T_{W}^{2}}B_{\mu }\right) ,  \notag \\
Z_{\mu }^{\prime } &=&-\sqrt{1-\beta ^{2}T_{W}^{2}}W_{\mu }^{8}+\beta
T_{W}B_{\mu },
\end{eqnarray}%
where the Weinberg angle is defined as 
\begin{equation}
S_{W}=\sin \theta _{W}=\frac{g^{\prime }}{\sqrt{g^{2}+\left( 1+\beta
^{2}\right) g^{\prime 2}}},\quad T_{W}=\tan \theta _{W}=\frac{g^{\prime }}{%
\sqrt{g^{2}+\beta ^{2}g^{\prime 2}}}
\end{equation}%
and $g,$ $g^{\prime }$ correspond to the coupling constants of the  $%
SU(3)_{L}$ and $U(1)_{X}$ groups, respectively. Further, a small mixing angle
between the two neutral currents $Z_{\mu }$ and $Z_{\mu }^{\prime }$ appears
with the following mass eigenstates \cite{beta-arbitrary}

\begin{eqnarray}
Z_{1\mu } &=&Z_{\mu }C_{\theta }+Z_{\mu }^{\prime }S_{\theta };\quad Z_{2\mu
}=-Z_{\mu }S_{\theta }+Z_{\mu }^{\prime }C_{\theta };  \notag \\
\tan \theta &=&\frac{1}{\Lambda +\sqrt{\Lambda ^{2}+1}};\quad \Lambda =\frac{%
-2S_{W}C_{W}^{2}g^{\prime 2}\nu _{\chi }^{2}+\frac{3}{2}S_{W}T_{W}^{2}g^{2}%
\left( \nu _{\eta }^{2}+\nu _{\rho }^{2}\right) }{gg^{\prime }T_{W}^{2}\left[
3\beta S_{W}^{2}\left( \nu _{\eta }^{2}+\nu _{\rho }^{2}\right)
+C_{W}^{2}\left( \nu _{\eta }^{2}-\nu _{\rho }^{2}\right) \right] }.
\label{mix}
\end{eqnarray}

\section{Neutral currents\label{sec:neutral-currents}}

Using the fermionic content from table \ref{tab:espectro}, we obtain the
following neutral couplings in weak eigenstates \cite{family-dependence}

\begin{eqnarray}
\mathcal{L}^{NC} &=&\frac{g}{2C_{W}}\left\{ \sum_{j=1}^{3}\overline{\text{Q}%
_{j}}\gamma _{\mu }\left[ g_{v}^{Q_{j}}-g_{a}^{Q_{j}}\gamma _{5}\right] 
\text{Q}_{j}Z^{\mu }+\overline{\ell _{j}}\gamma _{\mu }\left[ g_{v}^{\ell
_{j}}-g_{a}^{\ell _{j}}\gamma _{5}\right] \ell _{j}Z^{\mu }\right.  \notag \\
&&+\overline{\ell _{j}}\gamma _{\mu }\left[ \overset{\sim }{g}_{v}^{\ell
_{j}}-\overset{\sim }{g}_{a}^{\ell _{j}}\gamma _{5}\right] \ell _{j}Z^{\mu
\prime }+\sum_{m=1}^{2}\overline{q_{m^{\ast }}}\gamma _{\mu }\left[ \overset{%
\sim }{g}_{v}^{q_{m}}-\overset{\sim }{g}_{a}^{q_{m}}\gamma _{5}\right]
q_{m^{\ast }}Z^{\mu \prime }  \notag \\
&&\left. +\overline{q_{3}}\gamma _{\mu }\left[ \overset{\sim }{g}%
_{v}^{q_{3}}-\overset{\sim }{g}_{a}^{q_{3}}\gamma _{5}\right] q_{3}Z^{\mu
\prime }\right\} ,  \label{lag-1}
\end{eqnarray}%
where Q$_{j}$ with $j=1,2,3$ has been written in a SM-like notation i.e. it
refers to triplets of quarks associated with the three generations of quarks
(SM does not make a difference in the family representations). The vector and
axial vector couplings are shown in Table \ref{EW-couplings} for each
component and for any value of $\beta $. The results from Table \ref%
{EW-couplings} are in agreement with ref. \cite{sher}, where we can
reproduce the couplings of the charged leptons $e_{j}$ for models $L_{1}$ and 
$L_{2},$ which correspond in our case to $\beta =1/\sqrt{3}$ and $\beta =-1/%
\sqrt{3}$, respectively, and where the coupling constants are related as $%
\widetilde{g}_{v,a}=\frac{\mp S_{W}}{6}C_{v,a}$ for each case. The results
are also in agreement with ref. \cite{cesio2} for $\beta =\sqrt{3},-1/\sqrt{3%
}$. The vector and axial vector couplings can be written in a short form as

\begin{table}[tbp]
\begin{center}
\begin{tabular}{|c|c|c|c|c|}
\hline
$Fermion$ & $g_{v}^{f}$ & $g_{a}^{f}$ & $\widetilde{g}_{v}^{f}$ & $%
\widetilde{g}_{a}^{f}$ \\ \hline\hline
$\nu _{j}$ & $\frac{1}{2}$ & $\frac{1}{2}$ & $\frac{-1+(1-\sqrt{3}\beta
)S_{W}^{2}}{2\sqrt{3}\sqrt{1-(1+\beta ^{2})S_{W}^{2}}}$ & $\frac{-1+(1-\sqrt{%
3}\beta )S_{W}^{2}}{2\sqrt{3}\sqrt{1-(1+\beta ^{2})S_{W}^{2}}}$ \\ \hline
$e_{j}$ & $-\frac{1}{2}+2S_{W}^{2}$ & $-\frac{1}{2}$ & $\frac{-1+(1-3\sqrt{3}%
\beta )S_{W}^{2}}{2\sqrt{3}\sqrt{1-(1+\beta ^{2})S_{W}^{2}}}$ & $\frac{-1+(1+%
\sqrt{3}\beta )S_{W}^{2}}{2\sqrt{3}\sqrt{1-(1+\beta ^{2})S_{W}^{2}}}$ \\ 
\hline
$E_{j}$ & $2Q_{1}S_{W}^{2}$ & $0$ & $\frac{1-(1+2\sqrt{3}\beta
Q_{1})S_{W}^{2}}{\sqrt{3}\sqrt{1-(1+\beta ^{2})S_{W}^{2}}}$ & $\frac{%
1-S_{W}^{2}}{\sqrt{3}\sqrt{1-(1+\beta ^{2})S_{W}^{2}}}$ \\ \hline
$d_{m^{\ast }}$ & $-\frac{1}{2}+\frac{2}{3}S_{W}^{2}$ & $-\frac{1}{2}$ & $%
\frac{3-(3+\sqrt{3}\beta )S_{W}^{2}}{6\sqrt{3}\sqrt{1-(1+\beta ^{2})S_{W}^{2}%
}}$ & $\frac{1-(1-\sqrt{3}\beta )S_{W}^{2}}{2\sqrt{3}\sqrt{1-(1+\beta
^{2})S_{W}^{2}}}$ \\ \hline
$u_{m^{\ast }}$ & $\frac{1}{2}-\frac{4}{3}S_{W}^{2}$ & $\frac{1}{2}$ & $%
\frac{3-(3-5\sqrt{3}\beta )S_{W}^{2}}{6\sqrt{3}\sqrt{1-(1+\beta
^{2})S_{W}^{2}}}$ & $\frac{1-(1+\sqrt{3}\beta )S_{W}^{2}}{2\sqrt{3}\sqrt{%
1-(1+\beta ^{2})S_{W}^{2}}}$ \\ \hline
$J_{m^{\ast }}$ & $-2Q_{J_{m}}S_{W}^{2}$ & $0$ & $\frac{-1+(1+2\sqrt{3}\beta
Q_{J_{m}})S_{W}^{2}}{\sqrt{3}\sqrt{1-(1+\beta ^{2})S_{W}^{2}}}$ & $\frac{%
-1+S_{W}^{2}}{\sqrt{3}\sqrt{1-(1+\beta ^{2})S_{W}^{2}}}$ \\ \hline
$u_{3}$ & $\frac{1}{2}-\frac{4}{3}S_{W}^{2}$ & $\frac{1}{2}$ & $\frac{-3+(3+5%
\sqrt{3}\beta )S_{W}^{2}}{6\sqrt{3}\sqrt{1-(1+\beta ^{2})S_{W}^{2}}}$ & $%
\frac{-1+(1-\sqrt{3}\beta )S_{W}^{2}}{2\sqrt{3}\sqrt{1-(1+\beta
^{2})S_{W}^{2}}}$ \\ \hline
$d_{3}$ & $-\frac{1}{2}+\frac{2}{3}S_{W}^{2}$ & $-\frac{1}{2}$ & $\frac{%
-3+(3-\sqrt{3}\beta )S_{W}^{2}}{6\sqrt{3}\sqrt{1-(1+\beta ^{2})S_{W}^{2}}}$
& $\frac{-1+(1+\sqrt{3}\beta )S_{W}^{2}}{2\sqrt{3}\sqrt{1-(1+\beta
^{2})S_{W}^{2}}}$ \\ \hline
$J_{3}$ & $-2Q_{J_{3}}S_{W}^{2}$ & $0$ & $\frac{1-(1-2\sqrt{3}\beta
Q_{J_{3}})S_{W}^{2}}{\sqrt{3}\sqrt{1-(1+\beta ^{2})S_{W}^{2}}}$ & $\frac{%
1-S_{W}^{2}}{\sqrt{3}\sqrt{1-(1+\beta ^{2})S_{W}^{2}}}$ \\ \hline
\end{tabular}%
\end{center}
\caption{\textit{Vector and axial vector couplings of fermions and massive
neutral bosons} ($Z$,$Z^{\prime }$).}
\label{EW-couplings}
\end{table}

\begin{eqnarray}
g_{v}^{f} &=&T_{3}-2Q_{f}S_{W}^{2},  \notag \\
g_{a}^{f} &=&T_{3}  \notag \\
\overset{\sim }{g}_{v,a}^{q_{m}} &=&\frac{g^{\prime }C_{W}}{gT_{W}}\left[
T_{8}+\beta Q_{q_{m}}T_{W}^{2}\left( \frac{1}{2}\Lambda _{1}\pm 1\right) %
\right]  \notag \\
\overset{\sim }{g}_{v,a}^{q_{3}} &=&\frac{g^{\prime }C_{W}}{gT_{W}}\left[
-T_{8}+\beta Q_{q_{3}}T_{W}^{2}\left( \frac{1}{2}\Lambda _{2}\pm 1\right) %
\right]  \notag \\
\overset{\sim }{g}_{v,a}^{\ell _{j}} &=&\frac{g^{\prime }C_{W}}{gT_{W}}\left[
-T_{8}-\beta T_{W}^{2}\left( \frac{1}{2}\Lambda _{3}\mp Q_{\ell _{j}}\right) %
\right] ,  \label{coup0}
\end{eqnarray}%
where $f=$Q$_{j},$ $\ell _{j}$ in the first line and $Q_{f}$ the electric
charges. The Gell-Mann matrices $T_{3}$ $=\frac{1}{2}diag(1,-1,0)$ and $%
T_{8}=\frac{1}{2\sqrt{3}}diag(1,1,-2)$ are introduced in the notation. We
also define $\Lambda _{1}=diag(-1,\frac{1}{2},2)$ and $\Lambda _{2}=diag(%
\frac{1}{2},-1,2).$ Finally, $\ell _{j}$ denote the leptonic triplets with $%
\Lambda _{3}=diag(1,1,2Q_{1})$ and $Q_{1}$ defined as the electric charge of
the exotic leptons $E_{j}$ in table \ref{tab:espectro}. It is noted that $%
g_{v,a}^{f}$ are the same as the SM definitions and $\overset{\sim }{g}%
_{v,a}^{f}$ are $\beta $-dependent couplings of $Z_{\mu }^{\prime }$ (i.e.
model dependent). The couplings of $Z$ and all the couplings of leptons are
equal for $j=1,2,3$, so that these terms are universal and independent from
the representations of table \ref{tab:combination}. On the other hand, the
couplings of the additional gauge boson ($Z_{\mu }^{\prime }$) with the two
former families are different from the ones involving the third family. This
is because the third family transforms differently as remarked in
table \ref{tab:espectro}. Consequently, these terms depend from the
representation $A,$ $B$ or $C.$

The Lagrangian can be turned to a weak basis $U^{0}=\left(
u^{0},c^{0},t^{0}\right) ^{T},$ $D^{0}=\left( d^{0},s^{0},b^{0}\right) ^{T},$
$N^{0}=\left( \nu _{e}^{0},\nu _{\mu }^{0},\nu _{\tau }^{0}\right) ^{T},$ $%
E^{0}=\left( e^{0},\mu ^{0},\tau ^{0}\right) ^{T},$ where the exotic
fermions $J_{j}$ and $E_{j}$ have been omitted. In addition, the neutral
couplings can be written in terms of the mixing angle between $Z_{\mu }$ and 
$Z_{\mu }^{\prime }$ given by Eq. (\ref{mix}), where $Z_{1\mu }$ is the
SM-like neutral boson and $Z_{2\mu }$ the exotic ones. Taking a very small
angle, we can do C$_{\theta }\simeq 1$ so that the Lagrangian from Eq. (\ref%
{lag-1}) becomes 
\begin{eqnarray}
\mathcal{L}^{NC} &=&\frac{g}{2C_{W}}\left\{ \overline{U^{0}}\gamma _{\mu }%
\left[ G_{v}^{U(r)}-G_{a}^{U(r)}\gamma _{5}\right] U^{0}Z_{1}^{\mu }+%
\overline{D^{0}}\gamma _{\mu }\left[ G_{v}^{D(r)}-G_{a}^{D(r)}\gamma _{5}%
\right] D^{0}Z_{1}^{\mu }\right.  \notag \\
&&+\overline{N^{0}}\gamma _{\mu }\left[ G_{v}^{N}-G_{a}^{N}\gamma _{5}\right]
N^{0}Z_{1}^{\mu }+\overline{E^{0}}\gamma _{\mu }\left[ G_{v}^{E}-G_{a}^{E}%
\gamma _{5}\right] E^{0}Z_{1}^{\mu }  \notag \\
&&+\overline{U^{0}}\gamma _{\mu }\left[ \widetilde{G}_{v}^{U(r)}-\widetilde{G%
}_{a}^{U(r)}\gamma _{5}\right] U^{0}Z_{2}^{\mu }+\overline{D^{0}}\gamma
_{\mu }\left[ \widetilde{G}_{v}^{D(r)}-\widetilde{G}_{a}^{D(r)}\gamma _{5}%
\right] D^{0}Z_{2}^{\mu }  \notag \\
&&\left. +\overline{N^{0}}\gamma _{\mu }\left[ \widetilde{G}_{v}^{N}-%
\widetilde{G}_{a}^{N}\gamma _{5}\right] N^{0}Z_{2}^{\mu }+\overline{E^{0}}%
\gamma _{\mu }\left[ \widetilde{G}_{v}^{E}-\widetilde{G}_{a}^{E}\gamma _{5}%
\right] E^{0}Z_{2}^{\mu }\right\} ,  \label{lag-3}
\end{eqnarray}%
where the couplings associated with $Z_{1\mu }$ are

\begin{equation}
G_{v,a}^{f(r)}=g_{v,a}^{f}+\delta g_{v,a}^{f(r)};\qquad \delta
g_{v,a}^{f(r)}=\overset{\sim }{g}_{v,a}^{f(r)}S_{\theta },  \label{coup2}
\end{equation}%
and the couplings associated with $Z_{2\mu }$ are

\begin{equation}
\overset{\sim }{G}_{v,a}^{f(r)}=\overset{\sim }{g}_{v,a}^{f(r)}-\delta 
\overset{\sim }{g}_{v,a}^{f};\qquad \delta \overset{\sim }{g}%
_{v,a}^{f}=g_{v,a}^{f}S_{\theta },  \label{coup3}
\end{equation}%
where the label $(r)$ in $\widetilde{g}^{U,D(r)}$ refers to any of the
realizations $r=($ $A,$ $B$, $C)$ from table \ref{tab:combination}$.$ The $%
\beta -$dependent couplings for leptons from Eq. (\ref{coup0}) becomes

\begin{equation}
\overset{\sim }{g}_{v,a}^{\ell }=\frac{g^{\prime }C_{W}}{2gT_{W}}\left[ 
\frac{-1}{\sqrt{3}}-\beta T_{W}^{2}\pm 2Q_{\ell }\beta T_{W}^{2}\right] ,
\label{coup1}
\end{equation}

\noindent while for the quark couplings we get

\begin{equation}
\overset{\sim }{g}_{v,a}^{q(r)}=\frac{g^{\prime }C_{W}}{2gT_{W}}%
K^{(r)\dagger }\left[ \left( 
\begin{array}{ccc}
\frac{1}{\sqrt{3}}+\frac{\beta T_{W}^{2}}{3} &  &  \\ 
& \frac{1}{\sqrt{3}}+\frac{\beta T_{W}^{2}}{3} &  \\ 
&  & -\frac{1}{\sqrt{3}}+\frac{\beta T_{W}^{2}}{3}%
\end{array}%
\right) \pm 2Q_{q}\beta T_{W}^{2}\right] K^{(r)},  \label{coup1a}
\end{equation}%
with $q=U^{0},D^{0},$ and \noindent where we define for each representation
from table \ref{tab:combination}

\begin{equation}
K^{(A)}=I,\text{ }K^{(B)}=\left( 
\begin{array}{ccc}
1 & 0 & 0 \\ 
0 & 0 & 1 \\ 
0 & 1 & 0%
\end{array}%
\right) ,\text{ }K^{(C)}=\left( 
\begin{array}{ccc}
0 & 1 & 0 \\ 
0 & 0 & 1 \\ 
1 & 0 & 0%
\end{array}%
\right) .  \label{rep-rotation}
\end{equation}

We will consider linear combinations among the three families to obtain
couplings in mass eigenstates 
\begin{equation}
f^{0}=R_{f}f,  \label{rotation}
\end{equation}

\noindent where $f$ denotes the fermions in mass eigenstates, $f^{0}$ in
weak eigenstates and $R_{f}$ the rotation matrix that diagonalize the Yukawa
mass terms. Thus, we can write the Eq. (\ref{lag-3}) in mass eigenstates as

\begin{eqnarray}
\mathcal{L}^{NC} &=&\frac{g}{2C_{W}}\left\{ \left[ \overline{U}\gamma _{\mu
}\left( \mathfrak{G}_{v}^{U(r)}-\mathfrak{G}_{a}^{U(r)}\gamma _{5}\right) U+%
\overline{D}\gamma _{\mu }\left( \mathfrak{G}_{v}^{D(r)}-\mathfrak{G}%
_{a}^{D(r)}\gamma _{5}\right) D\right. \right.  \notag \\
&&+\left. \overline{N}\gamma _{\mu }\left( G_{v}^{N}-G_{a}^{N}\gamma
_{5}\right) N+\overline{E}\gamma _{\mu }\left( G_{v}^{E}-G_{a}^{E}\gamma
_{5}\right) E\right] Z_{1}^{\mu }  \notag \\
&&+\left[ \overline{U}\gamma _{\mu }\left( \widetilde{\mathfrak{G}}%
_{v}^{U(r)}-\widetilde{\mathfrak{G}}_{a}^{U(r)}\gamma _{5}\right) U+%
\overline{D}\gamma _{\mu }\left( \widetilde{\mathfrak{G}}_{v}^{D(r)}-%
\widetilde{\mathfrak{G}}_{a}^{D(r)}\gamma _{5}\right) D\right.  \notag \\
&&+\left. \left. \overline{N}\gamma _{\mu }\left( \widetilde{G}_{v}^{N}-%
\widetilde{G}_{a}^{N}\gamma _{5}\right) N+\overline{E}\gamma _{\mu }\left( 
\widetilde{G}_{v}^{E}-\widetilde{G}_{a}^{E}\gamma _{5}\right) E\right]
Z_{2}^{\mu }\right\} ,  \label{lag-5}
\end{eqnarray}%
where the couplings of quarks depend on the rotation matrix. Taking into
account the definitions from Eqs. (\ref{coup2}) and (\ref{coup3}), the
vector and axial vectors for $Z_{1}$ in Eq. (\ref{lag-5}) take the form

\begin{eqnarray}
R_{\ell }^{\dagger }G_{v,a}^{\ell }R_{\ell } &=&R_{\ell }^{\dagger }\left(
g_{v,a}^{\ell }+\delta g_{v,a}^{\ell }\right) R_{\ell }=G_{v,a}^{\ell }I, 
\notag \\
R_{q}^{\dagger }G_{v,a}^{q(r)}R_{q} &=&R_{q}^{\dagger }\left(
g_{v,a}^{q}+\delta g_{v,a}^{q(r)}\right) R_{q}=g_{v,a}^{q}I+\delta \mathfrak{%
g}_{v,a}^{q(r)}=\mathfrak{G}_{v,a}^{q(r)},  \label{coup4-a}
\end{eqnarray}

\noindent while for $Z_{2}$

\begin{eqnarray}
R_{\ell }^{\dagger }\widetilde{G}_{v,a}^{\ell }R_{\ell } &=&R_{\ell
}^{\dagger }\left( \widetilde{g}_{v,a}^{\ell }+\delta \widetilde{g}%
_{v,a}^{\ell }\right) R_{\ell }=\widetilde{G}_{v,a}^{\ell }I,  \notag \\
R_{q}^{\dagger }\widetilde{G}_{v,a}^{q(r)}R_{q} &=&R_{q}^{\dagger }\left( 
\widetilde{g}_{v,a}^{q(r)}+\delta \widetilde{g}_{v,a}^{q}\right) R_{q}=%
\widetilde{\mathfrak{g}}_{v,a}^{q(r)}+\delta g_{v,a}^{q}I=\widetilde{%
\mathfrak{G}}_{v,a}^{q(r)}.  \label{coup4-b}
\end{eqnarray}

Due to the fact that $g_{v,a}^{\ell }$ (SM couplings) and $\widetilde{g}%
_{v,a}^{\ell }$ in Eq. (\ref{coup1}) are family independent, the neutral
couplings in mass eigenstates of leptons are the same as in weak
eigenstates, such as indicated by the first lines in Eqs. (\ref{coup4-a})
and (\ref{coup4-b}). However, we obtain flavor changing couplings in the
quark sector due to the family dependence shown by $\widetilde{g}%
_{v,a}^{q(r)}$ in Eq. (\ref{coup1a}), as indicated in the second lines in
Eqs (\ref{coup4-a}) and (\ref{coup4-b}), where

\begin{equation}
\delta \mathfrak{g}_{v,a}^{q(r)}=R_{q}^{\dagger }\delta g_{v,a}^{q(r)}R_{q}=%
\widetilde{\mathfrak{g}}_{v,a}^{q(r)}S_{\theta },  \label{new-shift}
\end{equation}%
and

\begin{equation}
\widetilde{\mathfrak{g}}_{v,a}^{q(r)}=R_{q}^{\dagger }\widetilde{g}%
_{v,a}^{q(r)}R_{q}.  \label{matrix-coup}
\end{equation}

For the calculation, we adopt an ansatz on the texture of the quark mass
matrices in agreement with the physical masses and mixing angles. The $%
SU(3)_{L}\otimes U(1)_{X}$ Lagrangian for the Yukawa interaction between
quarks is

\begin{eqnarray}
-\mathcal{L}_{Yuk} &=&\sum_{m=1}^{2}\overline{q_{m^{\ast }L}}\left[ \Gamma
_{\eta }^{m^{\ast }D}\eta D_{R}^{0}+\Gamma _{\rho }^{m^{\ast }U}\rho
U_{R}^{0}+\Gamma _{\chi }^{m^{\ast }J}\chi J_{m^{\ast }R}^{0}\right]  \notag
\\
&&+\overline{q_{3L}}\left[ \Gamma _{\rho }^{3D}\rho D_{R}^{0}+\Gamma _{\eta
}^{3U}\eta U_{R}^{0}+\Gamma _{\chi }^{3J}\chi J_{3R}^{0}\right] +h.c,
\label{yukawa-1}
\end{eqnarray}%
with $\eta ,\rho $ and $\chi $ the scalar triplets from Table \ref%
{tab:quince} and $\Gamma _{\phi }^{iQ}$ being the Yukawa interaction
matrices. Taking into account only the $SU(2)_{L}$ subdoublets (which lie in
the two upper components of the triplet), and omitting the couplings of $%
\chi $, the mass eigenstates of the scalar sector can be written as \cite%
{beta-arbitrary} 
\begin{eqnarray}
H &=&\left( 
\begin{array}{c}
\phi _{1}^{\mp } \\ 
h_{3}^{0}+\nu \mp i\phi _{3}^{0}%
\end{array}%
\right) =\rho S_{\beta }-\eta ^{\ast }C_{\beta },  \notag \\
\phi &=&\left( 
\begin{array}{c}
h_{2}^{\mp } \\ 
-h_{4}^{0}\mp ih_{1}^{0}%
\end{array}%
\right) =\rho C_{\beta }+\eta ^{\ast }S_{\beta },  \label{scalar-rotation}
\end{eqnarray}%
where $\eta ^{\ast }$ denotes the conjugate representation of $\eta ,$ $\tan
\beta =\nu _{\rho }/\nu _{\eta }$ and $\nu =\sqrt{\nu _{\rho }^{2}+\nu
_{\eta }^{2}}.$ Thus, after some algebraic manipulation, the neutral
couplings of the Yukawa Lagrangian can be written as

\begin{eqnarray}
-\mathcal{L}_{Yuk}^{(0)} &=&\left[ \overline{D_{L}^{0}}\left( M_{D}\right)
D_{R}^{0}+\overline{U_{L}^{0}}\left( M_{U}\right) U_{R}^{0}\right] \left( 1+%
\frac{h_{3}^{0}\mp i\phi _{3}^{0}}{\nu }\right)  \notag \\
&&+\left[ \overline{D_{L}^{0}}\left( \Gamma _{D}\right) D_{R}^{0}+\overline{%
U_{L}^{0}}\left( \Gamma _{U}\right) U_{R}^{0}\right] \left( h_{4}^{0}\pm
ih_{1}^{0}\right) +h.c,  \label{yukawa-2}
\end{eqnarray}%
where the fermion masses and Yukawa coupling matrices are given by

\begin{equation}
M=\nu \left( \Gamma _{1}C_{\beta }+\Gamma _{2}S_{\beta }\right) \text{ \quad
and\quad\ }\Gamma =\Gamma _{1}S_{\beta }-\Gamma _{2}C_{\beta },
\label{yukawa-matrix}
\end{equation}%
where $\Gamma _{1}=\Gamma _{\eta }$ and $\Gamma _{2}=\Gamma _{\rho }$. The
Lagrangian from Eq. (\ref{yukawa-2}) is equivalent to the two-Higgs-doublet
model (2HDM) Lagrangian \cite{zhou}, which exhibits FCNC due to the non-diagonal
components of $\Gamma .$ In particular, we take the structure of mass matrix
suggested in ref. \cite{texture}, which is written in the basis $\left(
u^{0},c^{0},t^{0}\right) $ or $\left( d^{0},s^{0},b^{0}\right) $ as

\begin{equation}
M_{q}=\left( 
\begin{array}{ccc}
0 & A_{q} & A_{q} \\ 
A_{q} & B_{q} & C_{q} \\ 
A_{q} & C_{q} & B_{q}%
\end{array}%
\right) .  \label{texture}
\end{equation}%
As studied in ref. \cite{matsuda}, there are two possible assignments for
the texture components that reproduce the physical mixing angles of the CKM
matrix, each one associated with up and down quarks. For up-type quarks, $%
A_{U}=\sqrt{\frac{m_{t}m_{u}}{2}},$ $B_{U}=(m_{t}+m_{c}-m_{u})/2$ and $%
C_{U}=(m_{t}-m_{c}-m_{u})/2$; for down-type quarks $A_{D}=\sqrt{\frac{%
m_{d}m_{s}}{2}},$ $B_{D}=(m_{b}+m_{s}-m_{d})/2$ and $%
C_{D}=-(m_{b}-m_{s}+m_{d})/2$. The above ansatz is diagonalized by the
following rotation matrices \cite{matsuda}

\begin{equation}
R_{D}=\left( 
\begin{array}{ccc}
c & s & 0 \\ 
-\frac{s}{\sqrt{2}} & \frac{c}{\sqrt{2}} & -\frac{1}{\sqrt{2}} \\ 
-\frac{s}{\sqrt{2}} & \frac{c}{\sqrt{2}} & \frac{1}{\sqrt{2}} \\ 
&  & 
\end{array}%
\right) ;\text{ }R_{U}=\left( 
\begin{array}{ccc}
c^{\prime } & 0 & s^{\prime } \\ 
-\frac{s^{\prime }}{\sqrt{2}} & -\frac{1}{\sqrt{2}} & \frac{c^{\prime }}{%
\sqrt{2}} \\ 
-\frac{s^{\prime }}{\sqrt{2}} & \frac{1}{\sqrt{2}} & \frac{c^{\prime }}{%
\sqrt{2}} \\ 
&  & 
\end{array}%
\right) ,  \label{rot-matrix}
\end{equation}%
with

\begin{eqnarray}
c &=&\sqrt{\frac{m_{s}}{m_{d}+m_{s}}};\qquad s=\sqrt{\frac{m_{d}}{m_{d}+m_{s}%
}};  \notag \\
c^{\prime } &=&\sqrt{\frac{m_{t}}{m_{t}+m_{u}}};\qquad s^{\prime }=\sqrt{%
\frac{m_{u}}{m_{t}+m_{u}}}.  \label{rot-coef}
\end{eqnarray}%
For the quark masses, we use the running mass at $M_{Z_{1}}$ scale given by
Eq. (\ref{quarks-mass}) in the appendix \ref{appendixAA}, which lead us to
the following values

\begin{equation}
c=0.976;\quad s=0.219;\quad c^{\prime }=0.999;\quad s^{\prime }=0.00359.
\label{texture-components}
\end{equation}

\section{Constraints on the $Z-Z^{\prime }$ mixing and $Z_{2}$ mass for $%
\protect\beta $ arbitrary\label{sec:z-pole}}

The couplings of the $Z_{1\mu }$ bosons in Eq. (\ref{lag-5}) have the same form as
the SM-neutral couplings, where the vector and axial vector couplings $%
g_{V,A}^{SM}$ are replaced by $\mathfrak{G}_{V,A}=g_{V,A}^{SM}I+\delta 
\mathfrak{g}_{V,A},$ and the matrices $\delta \mathfrak{g}_{V,A}$ (given by
eq. (\ref{new-shift})) are corrections due to the small $Z_{\mu }-Z_{\mu
}^{\prime }$ mixing angle $\theta $ in mass eigenstates$.$ For this reason
all the analytical parameters at the Z-pole have the same SM form but with
small correction factors given by $\mathfrak{G}_{V,A}$ that depend on the
family assignment. In the SM, the partial decay widths of $Z_{1}$ into
fermions $f\overline{f}$ is described by \cite{one, pitch}:

\begin{equation}
\Gamma _{f}^{SM}=\frac{N_{c}^{f}G_{f}M_{Z_{1}}^{3}}{6\sqrt{2}\pi }\rho _{f}%
\sqrt{1-\mu _{f}^{2}}\left[ \left( 1+\frac{\mu _{f}^{2}}{2}\right) \left(
g_{v}^{f}\right) ^{2}+\left( 1-\mu _{f}^{2}\right) \left( g_{a}^{f}\right)
^{2}\right] R_{QED}R_{QCD},  \label{partial-decay}
\end{equation}

\noindent where $N_{c}^{f}=1$, 3 for leptons and quarks, respectively. $%
R_{QED}=1+\delta _{QED}^{f}$ and $R_{QCD}=1+\frac{1}{2}\left(
N_{c}^{f}-1\right) \delta _{QCD}^{f}$ are QED and QCD corrections given by
Eq. (\ref{QCD}) in appendix \ref{appendixA}, and $\mu
_{f}^{2}=4m_{f}^{2}/M_{Z}^{2}$ considers kinematical corrections only
important for the $b$-quark. Universal electroweak corrections sensitive to
the top quark mass are taken into account in $\rho _{f}=1+\rho _{t}$ and in $%
g_{V}^{SM}$ which is written in terms of an effective Weinberg angle \cite%
{one}

\begin{equation}
\overline{S_{W}}^{2}=\left( 1+\frac{\rho _{t}}{T_{W}^{2}}\right) S_{W}^{2},
\label{effective-angle}
\end{equation}

\noindent with $\rho _{t}=3G_{f}m_{t}^{2}/8\sqrt{2}\pi ^{2}$. Nonuniversal
vertex corrections are also taken into account in the $Z_{1}\overline{b}b$
vertex with additional one-loop leading terms which leads to $\rho _{b}=1-%
\frac{1}{3}\rho _{t}$ and $\overline{S_{W}}^{2}=\left( 1+\frac{\rho _{t}}{%
T_{W}^{2}}+\frac{2\rho _{t}}{3}\right) S_{W}^{2}$ \cite{one, pitch}.

Table \ref{tab:observables} from appendix \ref{appendixAA} summarizes some
observables at the $Z$ resonance, with their experimental values from CERN
collider (LEP), SLAC Liner Collider (SLC) and data from atomic parity
violation \cite{one}, the SM predictions, and the expressions predicted by
331 models. We use $M_{Z_{1}}=91.1876$ $GeV$, $S_{W}^{2}=0.23113$, and for
the predicted SM partial decay given by (\ref{partial-decay}), we use the
values from Eq. (\ref{SM-partial-decay}) (see appendix \ref{appendixAA}).

The 331 predictions from table \ref{tab:observables} in appendix \ref%
{appendixAA} are expressed for the LEP Z-pole observables in terms of SM
values corrected by

\begin{eqnarray}
\delta _{Z} &=&\frac{\Gamma _{u}^{SM}}{\Gamma _{Z}^{SM}}(\delta _{u}+\delta
_{c})+\frac{\Gamma _{d}^{SM}}{\Gamma _{Z}^{SM}}(\delta _{d}+\delta _{s})+%
\frac{\Gamma _{b}^{SM}}{\Gamma _{Z}^{SM}}\delta _{b}+3\frac{\Gamma _{\nu
}^{SM}}{\Gamma _{Z}^{SM}}\delta _{\nu }+3\frac{\Gamma _{e}^{SM}}{\Gamma
_{Z}^{SM}}\delta _{\ell };  \notag \\
\delta _{had} &=&R_{c}^{SM}(\delta _{u}+\delta _{c})+R_{b}^{SM}\delta _{b}+%
\frac{\Gamma _{d}^{SM}}{\Gamma _{had}^{SM}}(\delta _{d}+\delta _{s});  \notag
\\
\delta _{\sigma } &=&\delta _{had}+\delta _{\ell }-2\delta _{Z};  \notag \\
\delta A_{f} &=&\frac{\delta \mathfrak{g}_{V}^{ff}}{g_{V}^{f}}+\frac{\delta 
\mathfrak{g}_{A}^{ff}}{g_{A}^{f}}-\delta _{f},  \label{shift1}
\end{eqnarray}

\noindent where for the light fermions

\begin{equation}
\delta _{f}=\frac{2g_{v}^{f}\delta \mathfrak{g}_{v}^{ff}+2g_{a}^{f}\delta 
\mathfrak{g}_{a}^{ff}}{\left( g_{v}^{f}\right) ^{2}+\left( g_{a}^{f}\right)
^{2}},  \label{shift2}
\end{equation}

\noindent while for the $b$-quark

\begin{equation}
\delta _{b}=\frac{\left( 3-\beta _{K}^{2}\right) g_{v}^{b}\delta \mathfrak{g}%
_{v}^{bb}+2\beta _{K}^{2}g_{a}^{b}\delta \mathfrak{g}_{a}^{bb}}{\left( \frac{%
3-\beta _{K}^{2}}{2}\right) \left( g_{v}^{b}\right) ^{2}+\beta
_{K}^{2}\left( g_{a}^{b}\right) ^{2}}.  \label{shift3}
\end{equation}

\noindent The notation $\delta \mathfrak{g}_{v,a}^{ff}$ refers to the
diagonal part of the matrix $\delta \mathfrak{g}_{v,a}$ in Eq. (\ref%
{new-shift})$.$ The above expressions are evaluated in terms of the
effective Weinberg angle from Eq. (\ref{effective-angle}).

The weak charge is written as

\begin{equation}
Q_{W}=Q_{W}^{SM}+\Delta Q_{W}=Q_{W}^{SM}\left( 1+\delta Q_{W}\right) ,
\label{weak}
\end{equation}%
where $\delta Q_{W}=\frac{\Delta Q_{W}}{Q_{W}^{SM}}$. The deviation $\Delta
Q_{W}$ is \cite{cesio} 
\begin{equation}
\Delta Q_{W}=\left[ \left( 1+4\frac{S_{W}^{4}}{1-2S_{W}^{2}}\right) Z-N%
\right] \Delta \rho _{M}+\Delta Q_{W}^{\prime },  \label{dev}
\end{equation}%
and $\Delta Q_{W}^{\prime }$ which contains new physics gives

\begin{eqnarray}
\Delta Q_{W}^{\prime } &=&-16\left[ \left( 2Z+N\right) \left( g_{A}^{e}%
\overset{\sim }{\mathfrak{g}}_{v}^{uu}+\overset{\sim }{g}_{a}^{e}g_{V}^{u}%
\right) +\left( Z+2N\right) \left( g_{a}^{e}\overset{\sim }{\mathfrak{g}}%
_{v}^{dd}+\overset{\sim }{g}_{a}^{e}g_{v}^{d}\right) \right] S_{\theta } 
\notag \\
&&-16\left[ \left( 2Z+N\right) \overset{\sim }{g}_{a}^{e}\overset{\sim }{%
\mathfrak{g}}_{v}^{uu}+\left( Z+2N\right) \overset{\sim }{g}_{a}^{e}\overset{%
\sim }{\mathfrak{g}}_{v}^{dd}\right] \frac{M_{Z_{1}}^{2}}{M_{Z_{2}}^{2}}.
\label{new}
\end{eqnarray}

For cesium, and for the first term in (\ref{dev}) we take the value $\left[ \left( 1+4%
\frac{S_{W}^{4}}{1-2S_{W}^{2}}\right) Z-N\right] \Delta \rho _{M}\simeq
-0.01 $ \cite{cesio2,cesio}. With the definitions of the couplings $\overset{%
\sim }{g}_{V,A}^{\ell }$ in eq. (\ref{coup1}) we can see that the new physics
contribution given by Eq. (\ref{new}) is $\beta $-dependent, so that the
precision measurements are sensitive to the type of 331 model according to
the value of $\beta $. This dependence will allow us to perform precision
adjustments to $\beta $. We get the same correction for the spectrums $A$ and 
$B$ due to the fact that the weak charge depends mostly on the up-down
quarks, and $A,B$-cases maintain the same representation for this family.

With the expressions for the Z-pole observables and the experimental data
shown in table \ref{tab:observables}, we perform a $\chi ^{2}$ fit for each
representation $A,B$ and $C$ at 95\% CL, where the free quantities $%
S_{\theta },$ $M_{Z_{2}}$ and $\beta $ can be constrained at the $Z_{1}$
peak. We assume a covariance matrix with elements $V_{ij}=\rho _{ij}\sigma
_{i}\sigma _{j}$ among the Z-pole observables$,$ $\rho $ the correlation
matrix and $\sigma $ the quadratic root of the experimental and SM errors.
The $\chi ^{2}$ statistic with three degrees of freedom (d.o.f) is defined
as \cite{one}

\begin{equation}
\chi ^{2}(S_{\theta },M_{Z_{2}},\beta )=\left[ \mathbf{y}-\mathbf{F}%
(S_{\theta },M_{Z_{2}},\beta )\right] ^{T}V^{-1}\left[ \mathbf{y}-\mathbf{F}%
(S_{\theta },M_{Z_{2}},\beta )\right] ,  \label{chi}
\end{equation}

\noindent where $\mathbf{y=\{}y_{i}\mathbf{\}}$ represent the 22
experimental observables from table \ref{tab:observables}, and $\mathbf{F}$
the corresponding 331 prediction. Table \ref{tab:correlation} from appendix %
\ref{appendixAA} display the symmetrical correlation matrices taken from
ref. \cite{LEP}.

\begin{figure}[tbph]
\centering \includegraphics[scale=0.8]{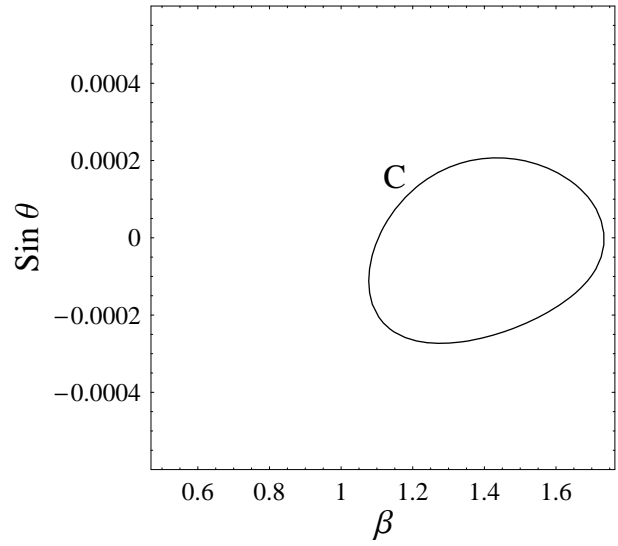}
\caption{\textit{The allowed region for $\sin\protect\theta$ vs $\protect%
\beta$ with $M_{Z_{2}}=1200$ GeV. C correspond to the assignment of family
from table 2. A and B assignments are excluded at this scale of $M_{Z_{2}}$.}
}
\label{figura1}
\end{figure}

\begin{figure}[tbph]
\centering \includegraphics[scale=0.8]{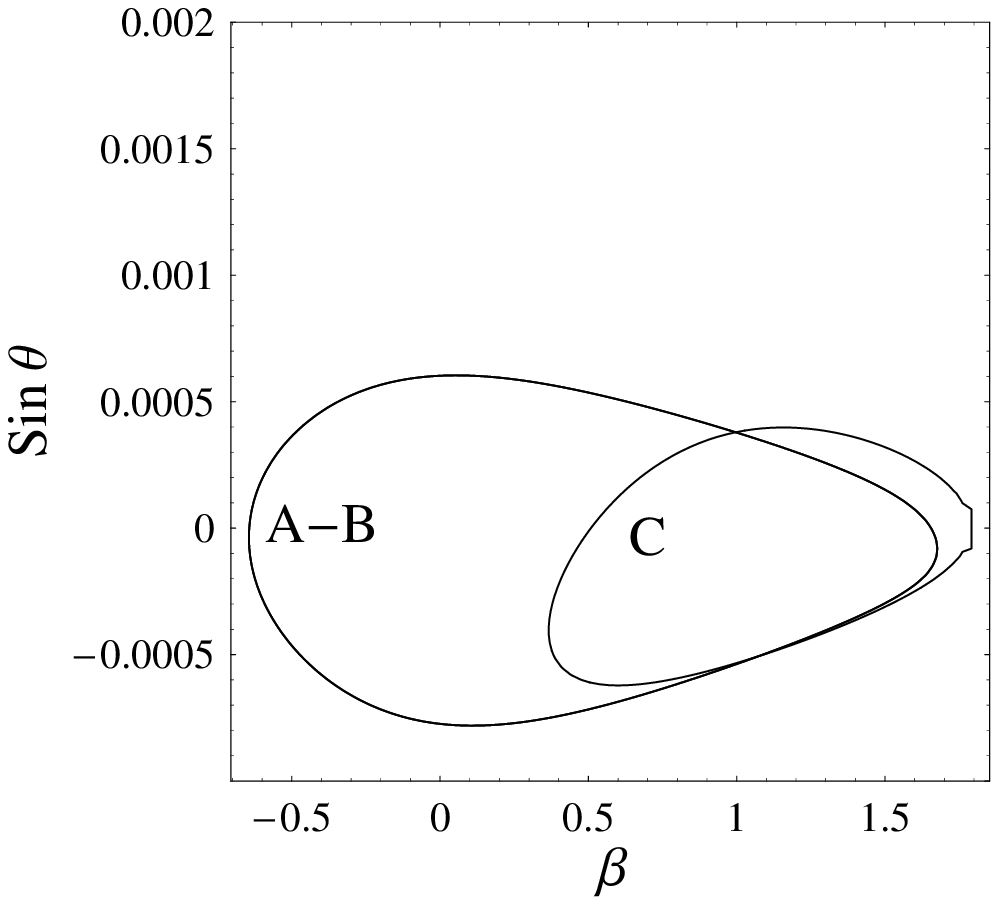}
\caption{\textit{The allowed region for $\sin\protect\theta$ vs $\protect%
\beta$ with $M_{Z_{2}}=1500$ GeV. A, B and C correspond to the assignment of
families from table 2.}}
\label{figura2}
\end{figure}

\begin{figure}[tbph]
\centering \includegraphics[scale=0.8]{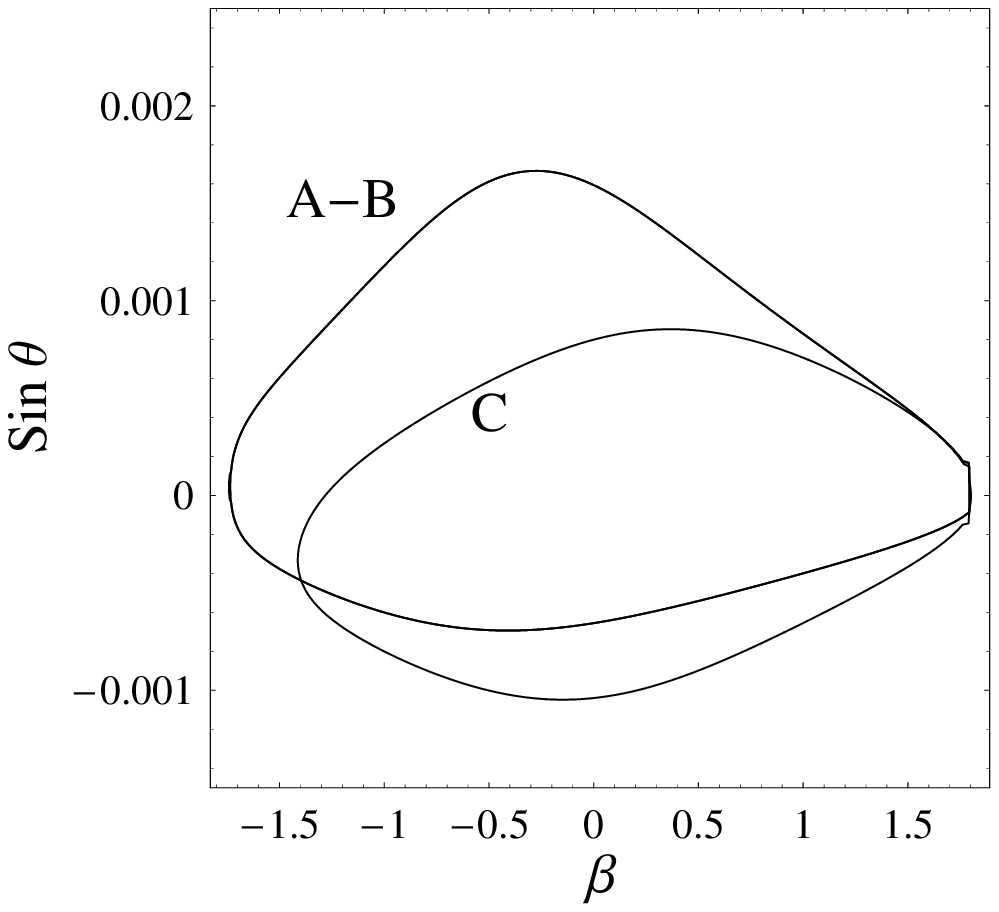}
\caption{\textit{The allowed region for $\sin\protect\theta$ vs $\protect%
\beta$ with $M_{Z_{2}}=4000$ GeV. A, B and C correspond to the assignment of
families from table 2}}
\label{figura3}
\end{figure}

\begin{figure}[tbph]
\centering \includegraphics[scale=0.8]{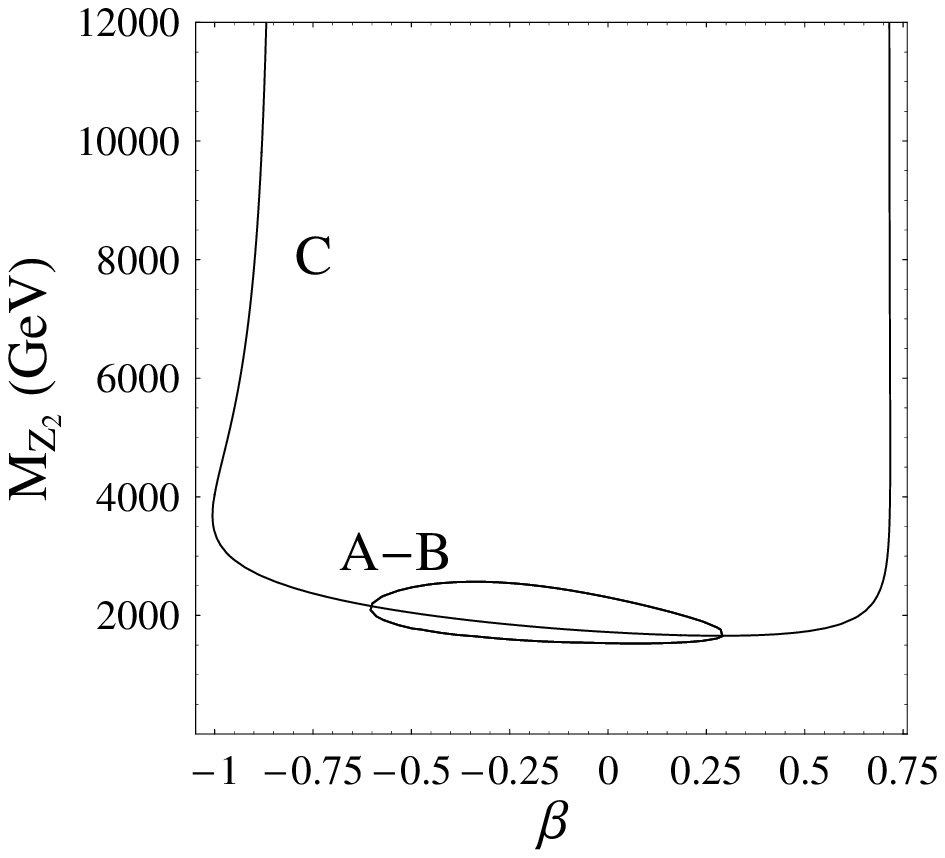}
\caption{\textit{The allowed region for $M_{Z_{2}}$ vs $\protect\beta$ with $%
\sin\protect\theta=-0.0008$. A, B and C correspond to the assignment of
families from table 2}}
\label{figura4}
\end{figure}

\begin{figure}[tbph]
\centering \includegraphics[scale=0.8]{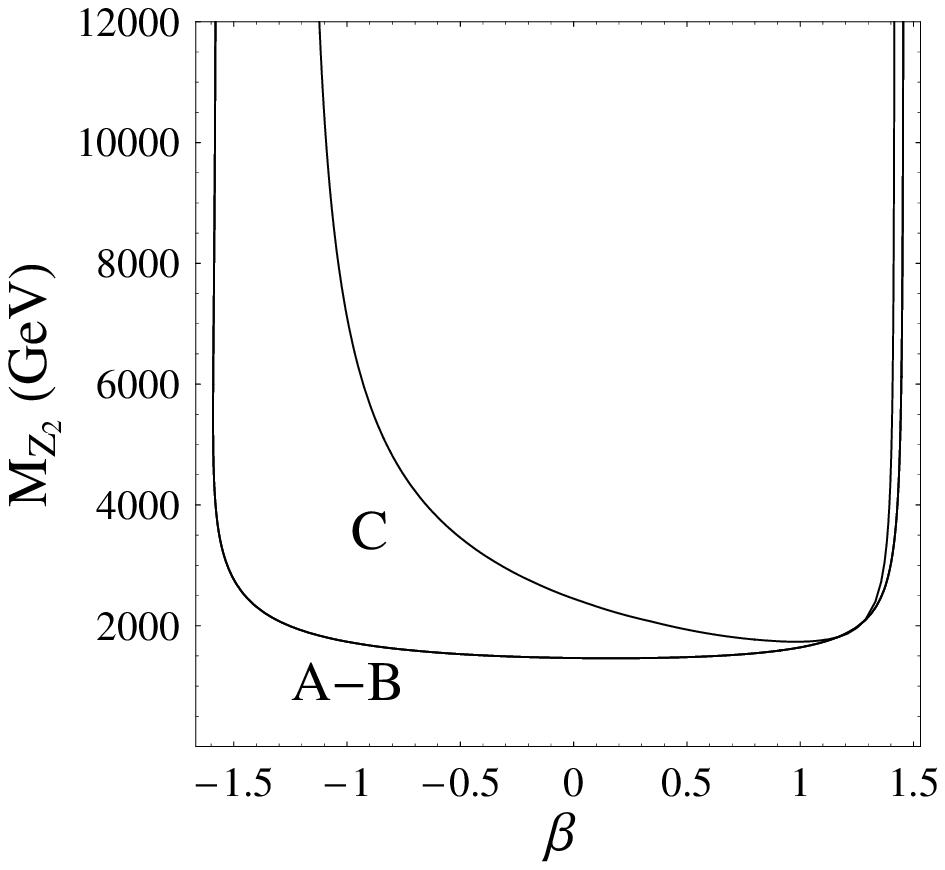}
\caption{\textit{The allowed region for $M_{Z_{2}}$ vs $\protect\beta$ with $%
\sin\protect\theta=0.0005$. A, B and C correspond to the assignment of
families from table 2}}
\label{figura5}
\end{figure}

\begin{figure}[tbph]
\centering \includegraphics[scale=0.8]{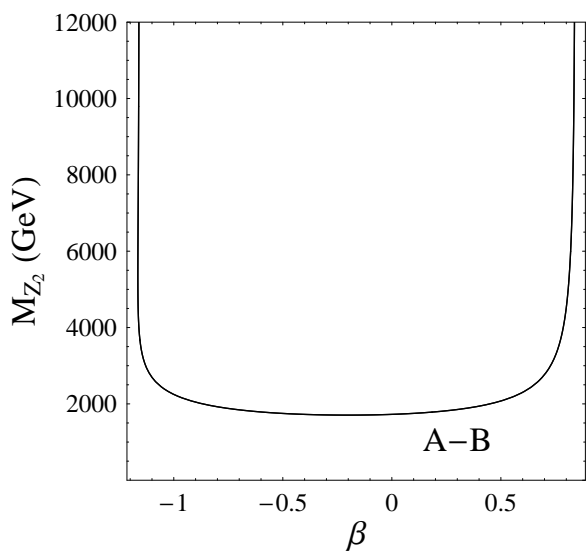}
\caption{\textit{The allowed region for $M_{Z_{2}}$ vs $\protect\beta$ with $%
\sin\protect\theta=0.001$. A and B correspond to the assignment of
families from table 2. The C assignment is excluded for this mixing angle.}}
\label{figura6}
\end{figure}

At three d.o.f, we get 3-dimensional allowed regions in the ($S_{\theta
},M_{Z_{2}},\beta $) space, which correspond to $\chi ^{2}\leq \chi _{\min
}^{2}+7.815,$ with $\chi _{\min }^{2}=16.98$ for $A$
and $B$ representations and $\chi _{\min }^{2}=19.27,$ for $C$. The plotted regions in Figs. %
\ref{figura1}-\ref{figura6} correspond to 2-dimensional cuts in the $%
S_{\theta }-\beta $ plane at $M_{Z_{2}}=1200,1500,4000$ $GeV,$ and in the $%
M_{Z_{2}}-\beta $ plane at $S_{\theta }=-0.0008,0.0005,0.001.$ The results
are summarized in tables \ref{tab:bound-1} and \ref{tab:bound-2}.

First of all, we find the best allowed region in the plane $S_{\theta
}-\beta $ for three different values of $M_{Z_{2}}$. The lowest bound of $%
M_{Z_{2}}$ that displays an allowed region is about $1200$ GeV, which
appears only for the C assignment such as Fig. \ref{figura1} shows. We can
see in the figure that models with negative values of $\beta $ are excluded,
including the usual models with $\beta =-\sqrt{3},-\frac{1}{\sqrt{3}}$. This
non-symmetrical behavior in the sign of $\beta $ is due to the fact that the
vector and axial couplings in Eqs. (\ref{coup1}) and (\ref{coup1a}) have a
lineal dependence with $\beta $, which causes different results according to
the sign. Figs. \ref{figura2} and \ref{figura3} display broader allowed
regions for $M_{Z_{2}}=1500$ and $4000$ GeV, respectively. Thus, the possible
331-models are highly restricted by low values of $M_{Z_{2}}$ (including the
exclusion of the main versions), but if the energy scale increases, new 331
versions are accessible. The models from literature are suitable for high
values of $Z_{2}-$mass. We also see that for small $Z_{2}-$mass, the bounds
associated with the mixing angle are very small ($\sim 10^{-4}$).

On the other hand, we obtain the regions in the plane $M_{Z_{2}}-\beta $ for
small values of $S_{\theta }$. Fig. \ref{figura4} shows regions for a
negative mixing angle ($S_{\theta }=-0.0008)$, where models with $\beta $ $<$
$0.75$ are favored. It is interesting to note that regions A and B display
thin bounds for $M_{Z_{2}}$ ($1500$ GeV$<M_{Z_{2}}<2500$ GeV). Figs. \ref%
{figura5} and \ref{figura6} show regions for positive mixing angles. In
particular, we can see in fig. \ref{figura6} that if $S_{\theta }=0.001$,
the C-family assignment does not display allowed region. Due to the fact
that the $A$ and $B$ spectrums present the same weak corrections, the allowed
regions coincide in all plots. We also see that the minima values for $%
M_{Z_{2}}$ are mostly extended in the positive values of $\beta ,$ although
not too far from zero. We emphasize that, although these results admit
continuous values of $\beta $, under some circumstances is possible to
obtain additional restrictions from basic principles that forbid some
specific values, as studied in ref. \cite{beta-arbitrary}.

\begin{table}[tbp]
\begin{center}
\begin{tabular}{c|c|c|c}
\hline
$M_{Z_{2}}$ (GeV) & Quarks Rep. & $\beta $ & $S_{\theta }$ ($\times 10^{-4}$)
\\ \hline\hline
1200 & Rep. $A-B$ & No Region & No Region \\ \cline{2-4}
& Rep. $C$ & $1.1\lesssim \beta \lesssim 1.75$ & $-3\leq S_{\theta }\leq 2$
\\ \hline\hline
1500 & Rep. $A-B$ & $-0.65\lesssim \beta \lesssim 1.7$ & $-8\leq S_{\theta
}\leq 6$ \\ \cline{2-4}
& Rep. $C$ & $0.35\lesssim \beta \lesssim 1.8$ & $-6\leq S_{\theta }\leq 4$
\\ \hline\hline
4000 & Rep. $A-B$ & $-1.75\lesssim \beta \lesssim 1.8$ & $-7\leq S_{\theta
}\leq 17$ \\ \cline{2-4}
& Rep. $C$ & $-1.4\lesssim \beta \lesssim 1.8$ & $-10\leq S_{\theta }\leq 8$
\\ \hline
\end{tabular}%
\end{center}
\caption{\textit{Bounds}\textit{\ for} $\protect\beta $\textit{\ and S}$_{%
\protect\theta }$ \textit{for three quark representations} \textit{at 95\%
CL and three} $Z_{2}$\textit{-mass} }
\label{tab:bound-1}
\end{table}

\begin{table}[tbp]
\begin{center}
\begin{tabular}{c|c|c|c}
\hline
$S_{\theta }$ ($\times 10^{-4}$) & Quarks Rep. & $\beta $ & $M_{Z_{2}}$ (GeV)
\\ \hline
$-8$ & Rep. $A-B$ & $-0.6\lesssim \beta \lesssim 0.3$ & $1500\lesssim
M_{Z_{2}}\lesssim 2500$ \\ \cline{2-4}
& Rep. $C$ & $-1\lesssim \beta \lesssim 0.75$ & $1500\lesssim M_{Z_{2}}$ \\ 
\hline\hline
5 & Rep. $A-B$ & $-1.6\lesssim \beta \lesssim 1.4$ & $1500\lesssim M_{Z_{2}}$
\\ \cline{2-4}
& Rep. $C$ & $-1.2\lesssim \beta \lesssim 1.5$ & $1700\lesssim M_{Z_{2}}$ \\ 
\hline\hline
10 & Rep. $A-B$ & $-1.15\lesssim \beta \lesssim 0.85$ & $1700\lesssim
M_{Z_{2}}$ \\ \cline{2-4}
& Rep. $C$ & No Region & No Region \\ \hline
\end{tabular}%
\end{center}
\caption{\textit{Bounds}\textit{\ for }$\protect\beta $\textit{\ and M}$%
_{Z_{2}}$ \textit{for three quark representations} \textit{at 95\% CL and
three mixing angle} $S_{\protect\theta }$ }
\label{tab:bound-2}
\end{table}

\section{The Z$_{2}$ decay for model with $\protect\beta =1/\protect\sqrt{3}$%
\label{Z'-decay}}

Since the oblique corrections are sensitive to heavy particles running into
the loop, we consider the one loop corrections to the $Z_{2}$ decay, taking
into account the exotic quarks $J$ from table \ref{tab:espectro}, where we
assume that the exotic spectrum is degenerated, and $%
m_{J_{1}}=m_{J_{2}}=m_{J_{3}}\gtrsim M_{Z_{2}}.$ In the $\overline{MS}$
scheme, all the infinite parts of the self-energies are subtracted by
properly adding divergent counterterms in the Lagrangian, while the finite
terms contribute to the corrections. These calculations are shown in the
appendix \ref{appendixA}, from where we get the following decay width 
\begin{equation}
\Gamma _{Z_{2}\rightarrow \overline{f}f}=\frac{g^{2}M_{Z_{2}}N_{c}}{48\pi
C_{W}^{2}}\sqrt{1-\mu _{f}^{\prime 2}}\left[ \left( 1+\frac{\mu _{f}^{\prime
2}}{2}\right) \left( \widetilde{\mathcal{G}}_{v}^{f}\right) ^{2}+\left(
1-\mu _{f}^{\prime 2}\right) \left( \widetilde{\mathcal{G}}_{a}^{f}\right)
^{2}\right] R_{QED}R_{QCD}.  \label{Z'-width}
\end{equation}

\noindent where $\mu _{f}^{\prime 2}=4m_{f}^{2}/M_{Z_{2}}^{2}$ takes into
account kinematical corrections only important for the top quark. The
corrections $R_{QED,QCD}$ are calculated at the $M_{Z_{2}}$ scale. The
effective couplings are

\begin{equation}
\widetilde{\mathcal{G}}_{v}^{f}=\widetilde{g}_{v}^{f}-\Delta \widetilde{g}%
_{v}^{f};\quad \widetilde{\mathcal{G}}_{a}^{f}=\widetilde{g}_{a}^{f}-\Delta 
\widetilde{g}_{a}^{f},  \label{new-vector-axial}
\end{equation}%
with $\widetilde{g}_{v,a}^{f}$ given by the Eqs. (\ref{coup1}) and (\ref%
{coup1a}). The effective radiative corrections evaluated at the $M_{Z_{2}}$
scale are given by

\begin{eqnarray}
\Delta \widetilde{g}_{v}^{f} &\approx &2S_{W}C_{W}Q_{f}\Pi _{Z_{2}\gamma
}(M_{z_{2}}^{2})+g_{v}^{f}\Pi _{Z_{2}Z_{1}}(M_{z_{2}}^{2})\left( 1+\frac{%
M_{Z_{1}}^{2}}{M_{Z_{2}}^{2}}\right) +\frac{1}{2}\widetilde{g}_{v}^{f}\Sigma
_{Z_{2}Z_{2}}^{\prime }(M_{z_{2}}^{2}),  \notag \\
\Delta \widetilde{g}_{a}^{f} &\approx &g_{a}^{f}\Pi
_{Z_{2}Z_{1}}(M_{z_{2}}^{2})\left( 1+\frac{M_{Z_{1}}^{2}}{M_{Z_{2}}^{2}}%
\right) +\frac{1}{2}\widetilde{g}_{a}^{f}\Sigma _{Z_{2}Z_{2}}^{\prime
}(M_{z_{2}}^{2}).  \label{effect-correction}
\end{eqnarray}

Using the definitions from Eqs. (\ref{new-vector-axial}) and (\ref%
{effect-correction}), the decay width can be written as (with $\left( \Delta 
\widetilde{g}\right) ^{2}\approx 0$)

\begin{equation}
\Gamma _{Z_{2}\rightarrow \overline{f}f}=\Gamma _{Z_{2}\rightarrow \overline{%
f}f}^{0}\left( 1-\Delta _{f}^{\prime }\right) ,  \label{Z'-decay2}
\end{equation}%

\noindent where the contribution at tree level is

\begin{equation}
\Gamma _{Z_{2}\rightarrow \overline{f}f}^{0}=\frac{g^{2}M_{Z_{2}}N_{c}}{%
48\pi C_{W}^{2}}\sqrt{1-\mu _{f}^{\prime 2}}\left[ \left( 1+\frac{\mu
_{f}^{\prime 2}}{2}\right) \left( \widetilde{g}_{v}^{f}\right) ^{2}+\left(
1-\mu _{f}^{\prime 2}\right) \left( \widetilde{g}_{a}^{f}\right) ^{2}\right]
R_{QED}R_{QCD},  \label{Z'-tree}
\end{equation}%

\noindent and

\begin{equation}
\Delta _{f}^{\prime }\approx \frac{2\left( \widetilde{g}_{v}^{f}\Delta 
\widetilde{g}_{v}^{f}+\widetilde{g}_{a}^{f}\Delta \widetilde{g}%
_{a}^{f}\right) }{\left( \widetilde{g}_{v}^{f}\right) ^{2}+\left( \widetilde{%
g}_{a}^{f}\right) ^{2}},  \label{ew-corrections}
\end{equation}%
that contains the oblique radiative corrections.

We calculate the decay widths taking into account the running coupling
constants at the $Z_{2}$ resonance. Using the following values at the $%
M_{Z_{1}}$ scale

\begin{eqnarray}
\alpha _{Y}^{-1}(M_{Z_{1}}) &=&98.36461\pm 0.06657;\quad \alpha
^{-1}(M_{Z_{1}})=127.934\pm 0.027;  \notag \\
\alpha _{2}^{-1}(M_{Z_{1}}) &=&29.56938\pm 0.00068;\quad \alpha
_{s}(M_{Z_{1}})=0.1187\pm 0.002;  \notag \\
S_{W}^{2}(M_{Z_{1}}) &=&0.23113\pm 0.00015,  \label{z-values}
\end{eqnarray}%
we get at the $M_{Z_{2}}\approx 1500$ GeV scale (see appendix \ref{appendixB}%
)

\begin{eqnarray}
\alpha _{Y}^{-1}(M_{Z_{2}}) &=&95.0867;\text{\qquad }\alpha
^{-1}(M_{Z_{2}})=125.993;  \notag \\
\alpha _{2}^{-1}(M_{Z_{2}}) &=&30.9060;\qquad \alpha _{s}(M_{Z_{2}})=0.0853;
\notag \\
S_{W}^{2}(M_{Z_{2}}) &=&0.2453,  \label{RCC}
\end{eqnarray}%
\bigskip while for the quark masses at $M_{Z_{2}}\approx 1500$ GeV, we get

\begin{eqnarray}
\overline{m_{u}}(M_{Z_{2}}) &=&0.00179\text{ GeV};\qquad \overline{m_{c}}%
(M_{Z_{2}})=0.537\;\text{GeV},  \notag \\
\overline{m_{t}}(M_{Z_{2}}) &=&150.73\text{ GeV;\qquad }\overline{m_{d}}%
(M_{Z_{2}})=0.00359\;\text{GeV};  \notag \\
\overline{m_{s}}(M_{Z_{2}}) &=&0.0717\;\text{GeV};\qquad \overline{m_{b}}%
(M_{Z_{2}})=2.45\;\text{GeV}.  \label{running-quarks-mass}
\end{eqnarray}

By comparing the data with radiative corrections to the decay $Z\rightarrow b%
\overline{b}$ carried out in ref \cite{Roberto-Sampayo}, exotic quarks with
a mass in the range $1.5-4$ TeV is found for the 331-bilepton model. Thus,
it is reasonable to estimate $m_{J_{j}}\approx 2$ $TeV.$ With the above
values and for $\beta =1/\sqrt{3},$ we obtain the widths shown in table \ref{tab:run-quark-tree} at
both tree and one-loop level from Eqs. (\ref{Z'-tree}) and (\ref{Z'-decay2}%
), respectively. In the leptonic sector, values independent on the family
representation (universal of family) are obtained in the two final rows in
table \ref{tab:run-quark-tree}, where the radiative corrections accounts for
about ($\Gamma ^{0}-\Gamma )/\Gamma ^{0}\approx 1.21$ and $1.13$ $\%$
deviation for charged and neutral leptons, respectively. In regard to the
quark widths, we obtain the family dependent decays shown in the table \ref{tab:run-quark-tree},
where relative deviations due to one-loop corrections are also calculated.

\begin{table}[tbp]
\begin{center}
\begin{tabular}{l|lll||lll||lll}
\hline
\multicolumn{4}{c||}{$\Gamma _{ff}^{0}$ (GeV)} & \multicolumn{3}{c||}{$%
\Gamma _{ff}$ (GeV)} & \multicolumn{3}{c}{$\frac{\Gamma ^{0}-\Gamma }{\Gamma
^{0}}\times 100$ (\%)} \\ \hline\hline
& A & \multicolumn{1}{|l}{B} & \multicolumn{1}{|l||}{C} & A & 
\multicolumn{1}{|l}{B} & \multicolumn{1}{|l||}{C} & A & \multicolumn{1}{|l}{B
} & \multicolumn{1}{|l}{C} \\ \hline
$u\overline{u}$ & 3.304 & \multicolumn{1}{|l}{3.304} & \multicolumn{1}{|l||}{
2.293} & 3.341 & \multicolumn{1}{|l}{3.341} & \multicolumn{1}{|l||}{2.317} & 
1.12 & \multicolumn{1}{|l}{1.12} & \multicolumn{1}{|l}{1.05} \\ \hline
$c\overline{c}$ & 3.304 & \multicolumn{1}{|l}{2.293} & \multicolumn{1}{|l||}{
3.304} & 3.341 & \multicolumn{1}{|l}{2.317} & \multicolumn{1}{|l||}{3.341} & 
1.12 & \multicolumn{1}{|l}{1.05} & \multicolumn{1}{|l}{1.12} \\ \hline
$t\overline{t}$ & 2.170 & \multicolumn{1}{|l}{3.271} & \multicolumn{1}{|l||}{
3.271} & 2.193 & \multicolumn{1}{|l}{3.307} & \multicolumn{1}{|l||}{3.307} & 
1.06 & \multicolumn{1}{|l}{1.1} & \multicolumn{1}{|l}{1.1} \\ \hline
$d\overline{d}$ & 2.974 & \multicolumn{1}{|l}{2.974} & \multicolumn{1}{|l||}{
1.963} & 3.006 & \multicolumn{1}{|l}{3.006} & \multicolumn{1}{|l||}{1.982} & 
1.07 & \multicolumn{1}{|l}{1.07} & \multicolumn{1}{|l}{0.97} \\ \hline
$s\overline{s}$ & 2.974 & \multicolumn{1}{|l}{1.963} & \multicolumn{1}{|l||}{
2.974} & 3.006 & \multicolumn{1}{|l}{1.982} & \multicolumn{1}{|l||}{3.006} & 
1.07 & \multicolumn{1}{|l}{0.97} & \multicolumn{1}{|l}{1.07} \\ \hline
$b\overline{b}$ & 1.963 & \multicolumn{1}{|l}{2.974} & \multicolumn{1}{|l||}{
2.974} & 1.982 & \multicolumn{1}{|l}{3.006} & \multicolumn{1}{|l||}{3.006} & 
0.97 & \multicolumn{1}{|l}{1.07} & \multicolumn{1}{|l}{1.07} \\ \hline\hline
$\ell ^{+}\ell ^{-}$ &  & 1.650 &  &  & 1.670 &  &  & 1.21 &  \\ \hline
$\nu \overline{\nu }$ &  & 1.327 &  &  & 1.342 &  &  & 1.13 &  \\ \hline
\end{tabular}%
\end{center}
\caption{\textit{Partial width of Z}$_{2}$ \textit{into fermions for each
representation A,B, and C. Leptons are universal of family. We compare the
relative deviations associated with the oblique radiative corrections.}}
\label{tab:run-quark-tree}
\end{table}

From the results in table \ref{tab:run-quark-tree}, it is possible to do a
rought estimative on the branching ratios. Assuming that only decays to SM
particles are allowed, we get  $Br(Z_{2}\rightarrow q\overline{q}%
)\sim 0.6$ for quarks, $Br(Z_{2}\rightarrow \ell ^{+}\overline{\ell
^{-}})\sim 0.2$ for charged leptons, and $Br(Z_{2}\rightarrow \nu \overline{\nu }%
)\sim 0.15$ for neutrinos. Comparing with other models \cite{tavares}, we get similar
results in the charged sector, but in the neutrino sector, we obtain a
branching about two order of magnitude bigger.

\section{Phenomenology on the FCNC\label{Z'-FCNC}}

In this section we introduce the flavor changing couplings from Eqs (\ref%
{coup4-a})-(\ref{matrix-coup}) in order to calculate the $Z_{1}$ and $Z_{2}$
decay widths. As in the above section, we will take the model $\beta =1/%
\sqrt{3}$

\subsection{The Z$_{1}$ decay with FCNC\label{Z-FCNC}}

The couplings from Eq (\ref{lag-5}) lead to the following partial width of $%
Z_{1}$ into fermions $f\overline{f^{\prime }}$

\begin{equation}
\Gamma _{ff^{\prime }}=\frac{N_{c}^{f}G_{f}M_{Z_{1}}^{3}}{6\sqrt{2}\pi }\rho
_{f}\left[ \left( \mathfrak{G}_{v}^{ff^{\prime }(r)}\right) ^{2}+\left( 
\mathfrak{G}_{a}^{ff^{\prime }(r)}\right) ^{2}\right] R_{QED}R_{QCD},
\label{partial-decay-2}
\end{equation}

\noindent where $\mathfrak{G}_{v,a}^{ff^{\prime }(r)}$ marks the $ff^{\prime
}$ component of the matrices given in Eq. (\ref{coup4-a}). We can see that
leptons contribute only for $\ell =\ell ^{\prime },$ while quarks may
exhibit FCNC due to the mixing angle and the non-universal couplings of $%
Z_{2}.$ Using the definitions from Eq. (\ref{coup4-a}) and (\ref{new-shift}%
), the FCNC contributions to the widths in Eq. (\ref{partial-decay-2}) can be written for
quarks as

\begin{equation}
\Gamma _{Z_{1}\rightarrow qq^{\prime }}=\frac{3G_{f}M_{Z_{1}}^{3}}{6\sqrt{2}%
\pi }\rho _{q}\left[ \left( \widetilde{\mathfrak{g}}_{v}^{qq^{\prime
}(r)}\right) ^{2}+\left( \widetilde{\mathfrak{g}}_{a}^{qq^{\prime
}(r)}\right) ^{2}\right] (S_{\theta })^{2}R_{QED}R_{QCD},
\label{Partial-decay-3}
\end{equation}%
where $\rho _{q}$ and $R_{QED,QCD}$ are
given by Eq. (\ref{partial-decay}). The above width gives the contribution
for processes with FCNC at tree level, where we can see that they are
suppressed by the small value $\left( S_{\theta }\right) ^{2}$. Table \ref%
{tab:ew-couplings} shows the values of the flavor changing electroweak
couplings $Z_{1}\overline{q}q^{\prime }$, where the ansatz from Eq. (\ref%
{rot-matrix}) is implemented. It is noted that the $Z_{1}$ decay into the
top quark is forbidden by kinematical conditions ($m_{t}>M_{Z_{1}}$). However,
it is possible to calculate the top quark width into the $Z_{1}$ boson and
light quarks $(u,c)$, obtaining

\begin{equation}
\Gamma _{t\rightarrow Z_{1}q}=\frac{\alpha m_{t}\left( 1-X_{Z}^{2}\right)
^{2}\left( 1+X_{Z}^{2}\right) }{16S_{W}^{2}C_{W}^{2}X_{Z}^{2}}\left[ \left( 
\widetilde{\mathfrak{g}}_{v}^{tq(r)}\right) ^{2}+\left( \widetilde{\mathfrak{%
g}}_{a}^{tq(r)}\right) ^{2}\right] (S_{\theta })^{2},  \label{top-decay}
\end{equation}%
with $X_{Z}=M_{Z_{1}}/m_{t}.$

\begin{table}[tbp]
\begin{center}
\begin{tabular}{cccc}
\hline
&  & $\widetilde{G}_{v,a}^{qq^{\prime }}$ &  \\ \hline\hline
& \multicolumn{1}{|c}{A} & \multicolumn{1}{|c}{B} & \multicolumn{1}{|c}{C}
\\ \hline\hline
$Z_{1}u\overline{c}$ & \multicolumn{1}{|c}{9.57$\times 10^{-4}S_{\theta }$}
& \multicolumn{1}{|c}{$-$9.57$\times 10^{-4}S_{\theta }$} & 
\multicolumn{1}{|c}{0} \\ \hline
$Z_{1}u\overline{t}$ & \multicolumn{1}{|c}{9.57$\times 10^{-4}S_{\theta }$}
& \multicolumn{1}{|c}{9.57$\times 10^{-4}S_{\theta }$} & \multicolumn{1}{|c}{%
$-1$9.15$\times 10^{-4}S_{\theta }$} \\ \hline
$Z_{1}c\overline{t}$ & \multicolumn{1}{|c}{$-0$.267$S_{\theta }$} & 
\multicolumn{1}{|c}{$0$.267$S_{\theta }$} & \multicolumn{1}{|c}{0} \\ \hline
$Z_{1}d\overline{s}$ & \multicolumn{1}{|c}{0.0569$S_{\theta }$} & 
\multicolumn{1}{|c}{0.0569$S_{\theta }$} & \multicolumn{1}{|c}{$-$0.114$%
S_{\theta }$} \\ \hline
$Z_{1}d\overline{b}$ & \multicolumn{1}{|c}{0.0583$S_{\theta }$} & 
\multicolumn{1}{|c}{$-$0.0583$S_{\theta }$} & \multicolumn{1}{|c}{0} \\ 
\hline
$Z_{1}s\overline{b}$ & \multicolumn{1}{|c}{$-0$.260$S_{\theta }$} & 
\multicolumn{1}{|c}{$0$.260$S_{\theta }$} & \multicolumn{1}{|c}{0} \\ \hline
\end{tabular}%
\end{center}
\caption{\textit{Vector and axial couplings of Z}$_{1}qq^{\prime }$ \textit{%
vertex for each representation A,B, and C. These results correspond to the
model }$\protect\beta =1/\protect\sqrt{3},$ \textit{where} $-0.0007$ $\leq
S_{\protect\theta }\leq 0.0005$ \textit{for A and B representations, and }$%
-0.0005$ $\leq S_{\protect\theta }\leq 0.0001$ \textit{for C representation. 
}}
\label{tab:ew-couplings}
\end{table}

Using the same values from section \ref{sec:z-pole} at $Z_{1}-$pole, we
obtain the FCNC widths shown in table \ref{tab:FCNC-Z-decay} as a fraction
of the quadratic mixing angle value $S_{\theta }^{2},$ and for each
representation $A,B$ and $C$.

\begin{table}[tbp]
\begin{center}
\begin{tabular}{c|c|c}
\hline
$\Gamma _{qq^{\prime }}/S_{\theta }^{2}$ (MeV) & A-B & C \\ \hline\hline
$Z_{1}\rightarrow uc$ & 1.917$\times 10^{-3}$ & 0 \\ \hline
$Z_{1}\rightarrow ds$ & 6.776 & 27.103 \\ \hline
$Z_{1}\rightarrow db$ & 7.116 & 0 \\ \hline
$Z_{1}\rightarrow sb$ & 141.715 & 0 \\ \hline\hline
$\Gamma _{Zq}/S_{\theta }^{2}$ (MeV) & A-B & C \\ \hline
$t\rightarrow Z_{1}u$ & 4.043$\times 10^{-3}$ & 16.172$\times 10^{-3}$ \\ 
\hline
$t\rightarrow Z_{1}c$ & 314.081 & 0 \\ \hline
\end{tabular}%
\end{center}
\caption{\textit{Partial width of Z}$_{1}$ \textit{into quarks and top quark
into Z}$_{1}$-\textit{quarks with FCNC for each representation A,B, and C,
as a fraction of the quadratic mixing angle value. This results correspond
to the model }$\protect\beta =1/\protect\sqrt{3},$ \textit{where} $-0.0007$ $%
\leq S_{\protect\theta }\leq 0.0005$ \textit{for A and B representations, and 
}$-0.0005$ $\leq S_{\protect\theta }\leq 0.0001$ \textit{for C
representation. }}
\label{tab:FCNC-Z-decay}
\end{table}

First, we note that the decays are highly constrained by the quadratic value
of the mixing angle. In particular, we can see from Fig. 2 that, for $\beta
=1/\sqrt{3},$ the bounds for the mixing angle in A-B cases are about $%
-0.0007\leq S_{\theta }\leq 0.0005$, from where the FCNC are suppressed by
the maximum factor $S_{\theta }^{2}=4.9\times 10^{-7}.$ Thus, the maxima
contributions to the decays $Z_{1}\rightarrow uc$ and $t\rightarrow Z_{1}u$
are of the order of $10^{-9}-10^{-10}$ MeV, while for the decays $%
Z_{1}\rightarrow sb$ and $t\rightarrow Z_{1}c$ are about $10^{-4}$ MeV. In
particular, we get for the top decay the branching $Br(t\rightarrow
Z_{1}c)\approx 10^{-7}$. For A-B representations, the decay of the top-quark
into the charm quark (i.e. $t\rightarrow Z_{1}c$) is about $10^{5}$ times
bigger than into the up quark. The source of this difference comes from
the texture structure in Eq. (\ref{texture}) and the rotation matrices in Eq
(\ref{rot-matrix}). A thorough check of Eq. (\ref{matrix-coup}), lead us to
the following proportions

\begin{eqnarray}
\widetilde{\mathfrak{g}}_{v,a}^{ut(A,B)} &\sim &s^{\prime }=\sqrt{\frac{m_{u}%
}{m_{t}+m_{u}}}=0.0036,  \notag \\
\widetilde{\mathfrak{g}}_{v,a}^{ct(A,B)} &\sim &c^{\prime }=\sqrt{\frac{m_{t}%
}{m_{t}+m_{u}}}=0.999,  \label{proportion}
\end{eqnarray}%
where we take the values from Eq. (\ref{quarks-mass}) in appendix \ref%
{appendixAA} at the $M_{Z_{1}}$ scale. Since the FCNC contributions depends
as the quadratic value of the couplings (see Eq. (\ref{Partial-decay-3})),
we obtain a contribution of $s^{\prime 2}\sim 10^{-5}$ and $c^{\prime 2}\sim
1$ for the decay into $u$ and $c,$ respectively. A similar result is
obtained for the down sector, where the width of $Z_{1}$ into $s\overline{b}$
quarks is $10^{2}$ times bigger than into $d\overline{b}(\overline{s})$
quarks. Thus, the hierarchical order found in the decay widths from table %
\ref{tab:FCNC-Z-decay} arises as the result of the mass hierarchical order $%
m_{u,d}\ll m_{t,b}.$ On the other hand, we see that representation $C$
suppress most of the flavor changing decays, which arises as a result of the
family structure exhibited by this representation in table \ref%
{tab:combination} when written in mass eigenstates through the rotation
matrices from Eq. (\ref{rot-matrix}).

\subsection{The Z$_{2}$ decay with FCNC}

Now, we consider the Lagrangian in mass eigenstates from Eq. (\ref{lag-5})
in the $S_{\theta }=0$ limit, where the corrections $\delta g_{V,A}$
dissapear. Thus, the Lagrangian that describes FCNC in Eq. (\ref{lag-5})
takes the form 
\begin{equation}
\mathcal{L}^{FCNC}=\frac{g}{2C_{W}}\left[ \overline{U}\gamma _{\mu }\left( 
\widetilde{\mathfrak{g}}_{v}^{U(r)}-\widetilde{\mathfrak{g}}%
_{a}^{U(r)}\gamma _{5}\right) U+\overline{D}\gamma _{\mu }\left( \widetilde{%
\mathfrak{g}}_{v}^{D(r)}-\widetilde{\mathfrak{g}}_{a}^{D(r)}\gamma
_{5}\right) D\right] Z_{2}^{\mu },  \label{lag-FCNC}
\end{equation}

\noindent with $\widetilde{\mathfrak{g}}_{v,a}$ defined by Eq. (\ref%
{matrix-coup}). We take the definitions of the texture structure from Eqs. (%
\ref{texture}) and (\ref{rot-matrix}), whose components are determined by
the values from Eq. (\ref{rot-coef}) but with quark masses given by (\ref%
{running-quarks-mass}) at $M_{Z_{2}}$ scale, which leads to

\begin{equation}
c=0.976;\quad s=0.218;\quad c^{\prime }=0.999;\quad s^{\prime }=0.00344.
\label{rot-coef-2}
\end{equation}%
We can see that the above components are very similar to the
values in Eq. (\ref{texture-components}) at the $Z_{1}$ scale. The values of the couplings
associated with the $Z_{2}qq^{\prime }$ vertices are given in Table \ref%
{ew-z2-couplings}. The width of $Z_{2}$ into different flavors of quarks $q%
\overline{q^{\prime }}$ gives

\begin{equation}
\Gamma _{Z_{2}\rightarrow qq^{\prime }}=\Gamma _{Z_{2}\rightarrow qq^{\prime
}}^{0}\left( 1-\Delta _{qq^{\prime }}^{\prime }\right) ,  \label{decay-FCNC}
\end{equation}%
where the tree-level contribution is

\begin{equation}
\Gamma _{Z_{2}\rightarrow qq^{\prime }}^{0}=\frac{g^{2}M_{Z_{2}}N_{c}}{48\pi
C_{W}^{2}}\left[ \left( \widetilde{\mathfrak{g}}_{v}^{qq^{\prime }}\right)
^{2}+\left( \widetilde{\mathfrak{g}}_{a}^{qq^{\prime }}\right) ^{2}\right] ,
\label{Z'-decay-tree}
\end{equation}%
and the radiative corrections due to the $Z_{2}-Z_{2}$ self-energy are
contained in

\begin{equation}
\Delta _{qq^{\prime }}=\frac{2\left( \widetilde{g}_{v}^{qq^{\prime }}\Delta 
\widetilde{g}_{v}^{qq^{\prime }}+\widetilde{g}_{a}^{qq^{\prime }}\Delta 
\widetilde{g}_{a}^{qq^{\prime }}\right) }{\left( \widetilde{g}%
_{v}^{qq^{\prime }}\right) ^{2}+\left( \widetilde{g}_{a}^{qq^{\prime
}}\right) ^{2}},  \label{FCNC-corrections}
\end{equation}%
with

\begin{equation}
\Delta \widetilde{g}_{v,a}^{qq^{\prime }}\approx \frac{1}{2}\widetilde{g}%
_{v,a}^{qq^{\prime }}\Sigma _{Z_{2}Z_{2}}^{\prime }(M_{Z_{2}}^{2}),
\label{FC-correction-couplings}
\end{equation}

\noindent where $\Sigma _{Z_{2}Z_{2}}^{\prime }$ is given by Eq. (\ref%
{derivative-Z'-Z'-self}) in Appendix \ref{appendixA}. We consider running coupling constants evaluated
at the $M_{Z_{2}}$ scale, which are given in Eq. (\ref{RCC}). In particular,
we take the model with $\beta =1/\sqrt{3},$ whose extra particles do not
present exotic charges, and has the lowest bound $M_{Z_{2}}\approx 1.5$ $TeV.
$ For the quarks running into the loop, we take again $m_{J_{j}}\approx 2$ $%
TeV$ \cite{Roberto-Sampayo}. The results are summarized in table (\ref%
{tab:FCNC-Zprima-decay}).

\begin{table}[tbp]
\begin{center}
\begin{tabular}{cccc}
\hline
&  & $\widetilde{g}_{v,a}^{qq^{\prime }}$ &  \\ \hline\hline
& \multicolumn{1}{|c}{A} & \multicolumn{1}{|c}{B} & \multicolumn{1}{|c}{C}
\\ \hline\hline
$Z_{2}u\overline{c}$ & \multicolumn{1}{|c}{9.15$\times 10^{-4}$} & 
\multicolumn{1}{|c}{$-$9.15$\times 10^{-4}$} & \multicolumn{1}{|c}{0} \\ 
\hline
$Z_{2}u\overline{t}$ & \multicolumn{1}{|c}{9.15$\times 10^{-4}$} & 
\multicolumn{1}{|c}{9.15$\times 10^{-4}$} & \multicolumn{1}{|c}{$-1$8.29$%
\times 10^{-4}$} \\ \hline
$Z_{2}c\overline{t}$ & \multicolumn{1}{|c}{$-0$.265} & \multicolumn{1}{|c}{$%
0 $.265} & \multicolumn{1}{|c}{0} \\ \hline
$Z_{2}d\overline{s}$ & \multicolumn{1}{|c}{0.0566} & \multicolumn{1}{|c}{
0.0566} & \multicolumn{1}{|c}{$-$0.113} \\ \hline
$Z_{2}d\overline{b}$ & \multicolumn{1}{|c}{0.0580} & \multicolumn{1}{|c}{$-$%
0.0580} & \multicolumn{1}{|c}{0} \\ \hline
$Z_{2}s\overline{b}$ & \multicolumn{1}{|c}{$-0$.259} & \multicolumn{1}{|c}{$%
0 $.259} & \multicolumn{1}{|c}{0} \\ \hline
\end{tabular}%
\end{center}
\caption{\textit{Vector and axial couplings of Z}$_{2}qq^{\prime }$ \textit{%
vertex for each representation A,B, and C. This results correspond to the
model }$\protect\beta =1/\protect\sqrt{3}$\textit{. }}
\label{ew-z2-couplings}
\end{table}

\begin{table}[tbp]
\begin{center}
\begin{tabular}{ccc}
\hline
$\Gamma _{qq^{\prime }}$ (MeV) & A-B & C \\ \hline\hline
$Z_{2}\rightarrow u\overline{c}$ & 0.0272 & 0 \\ \hline
$Z_{2}\rightarrow u\overline{t}$ & 0.0272 & 0.109 \\ \hline
$Z_{2}\rightarrow c\overline{t}$ & 2290.65 & 0 \\ \hline
$Z_{2}\rightarrow d\overline{s}$ & 104.15 & 416.59 \\ \hline
$Z_{2}\rightarrow d\overline{b}$ & 109.37 & 0 \\ \hline
$Z_{2}\rightarrow s\overline{b}$ & 2181.31 & 0 \\ \hline
\end{tabular}%
\end{center}
\caption{\textit{Partial width of Z}$_{2}$ \textit{into quarks with FCNC for
each representation A,B, and C. These values correspond to the model }$%
\protect\beta =1/\protect\sqrt{3}$ }
\label{tab:FCNC-Zprima-decay}
\end{table}
First of all, we note that these values are almost one order of magnitude
bigger than the fractions obtained in table \ref{tab:FCNC-Z-decay}
associated with the $Z_{1}$ boson decay. This behavior is due to the fact
that the FCNC widths as a fraction of the mixing angle $S_{\theta }^{2}$ in
Eqs. (\ref{Partial-decay-3}) and (\ref{top-decay}) depends on the factor ($%
\widetilde{\mathfrak{g}}_{v}$)$^{2}+$($\widetilde{\mathfrak{g}}_{a}$)$^{2},$
which is the same as the one written in Eq. (\ref{Z'-decay-tree}). The
differences between both cases come basically from the multiplicative
factors; in one case proportional to the $Z_{1}$ boson mass $%
M_{Z_{1}}=91.1876$ GeV, and in the other case to the $Z_{2}$ mass $%
M_{Z_{2}}\sim 1500$ GeV. In addition, the hierarchical order found in the
values from table \ref{tab:FCNC-Zprima-decay} is a direct consequence from
the dependence shown by Eq. (\ref{proportion}), where the $u\overline{c}(%
\overline{t})$ and $d\overline{s}(\overline{b})$ decays are lower in about $%
10^{-5}$ and $10^{-2}$ order of magnitude, respectively.

\section{Conclusions\label{conclusions}}

We found three different assignments of quarks into families in mass eigenstates.
Each assignment determines different weak couplings of the quarks to the
extra neutral current associated with $Z'$, which exhibits a small mixing angle with the SM-neutral boson $Z.$ The
Lagrangian of the Yukawa interaction is equivalent to the 2HDM, which
presents flavor changing neutral currents (FCNC) associated with the
couplings of the neutral scalars. In particular we adopt the ansatz shown in
Eq (\ref{texture}) proposed in ref. \cite{matsuda}. With this texture on the
matrices, we studied the constraints on the $Z-Z^{\prime }$ mixing and $%
Z_{2} $ mass for $\beta $ arbitrary, obtaining different allowed regions in
the $S_{\theta }-\beta $ and $M_{Z_{2}}-\beta $ planes for the LEP
parameters at the Z-pole. Through a $\chi ^{2}$ fit at the 95\% CL and 3
d.o.f, we found regions in the $S_{\theta }-\beta $ plane that display a
dependence in the family assignment for different values of $M_{Z_{2}}$
(figs. $1-3$). For the lowest value $M_{Z_{2}}=1200$ GeV, we found that only
those 331 models with $1.1\lesssim \beta \lesssim 1.75$ and quarks families
in the C-representation, yield a possible region with small mixing angles ($%
\sim 10^{-4}$). The possibilities of 331-models grow as $M_{Z_{2}}$ grows,
exhibiting broader regions for the mixing angle. For the $M_{Z_{2}}-\beta $
plots (figs. $4-6$), we also found model and family restrictions according
to the mixing angle. In this case, the $\beta $-bound grows when the mixing
angle decreases near zero. This behavior seen in the three figures is in
agreement with the results from figs. $1-3$, where the bounds for $\beta $
acquire their maxima values around $S_{\theta }=0$. The Pleitez and Long
models ($\beta =-\sqrt{3},-\frac{1}{\sqrt{3}}$, respectively) are excluded
for low values of $M_{Z_{2}}$ ($<1500$ GeV). In fact, we found that the
lowest bounds in the $M_{Z_{2}}$ value are found in regions with $\beta
\gtrsim 0.$

The $Z_{2}$ decay at $\sqrt{s}\approx M_{Z_{2}}$ was also studied, where the 
$Z-Z^{\prime }$ mixing is highly suppressed. We take the model $\beta =1/%
\sqrt{3}$ for the numerical calculations$,$ which holds a typical bound $%
M_{Z_{2}}\approx 1500$ GeV and does not exhibit exotic charges in the
spectrum. The decay widths were evaluated for each family representation and
taking into account oblique radiative corrections associated with the heaviest
quarks $J_{1,2,3}$ from table \ref{tab:espectro} and considering $%
M_{Z_{2}}<m_{J_{j}}\approx 2000$ GeV. We also considered the running
coupling constants at the $M_{Z_{2}}$ scale, obtaining decay widths with
values between 1.34 GeV and 3.34 GeV (see table \ref{tab:run-quark-tree}).
The tree level contributions were also calculated. The radiative corrections
account for about $1\%$ deviations. These radiative corrections are
sensitive to the mass $m_{J_{j}}$ of the quarks running into the loop. For
instance, the table \ref{tab:run-quark-tree2} shows the loop contributions
in the scenery with $m_{J_{j}}\approx 4000$ GeV \cite{Roberto-Sampayo},
where we can see that the radiative corrections account for about $5\%$
deviations.

\begin{table}[tbp]
\begin{center}
\begin{tabular}{l|lll||lll}
\hline
\multicolumn{4}{c||}{$\Gamma _{ff}$ (GeV)} & \multicolumn{3}{c}{$\frac{%
\Gamma ^{0}-\Gamma }{\Gamma ^{0}}\times 100$ (\%)} \\ \hline\hline
& A & \multicolumn{1}{|l}{B} & \multicolumn{1}{|l||}{C} & A & 
\multicolumn{1}{|l}{B} & \multicolumn{1}{|l}{C} \\ \hline
$u\overline{u}$ & 3.470 & \multicolumn{1}{|l}{3.470} & \multicolumn{1}{|l||}{
2.404} & 4.74 & \multicolumn{1}{|l}{4.74} & \multicolumn{1}{|l}{4.57} \\ 
\hline
$c\overline{c}$ & 3.470 & \multicolumn{1}{|l}{2.404} & \multicolumn{1}{|l||}{
3.470} & 4.74 & \multicolumn{1}{|l}{4.57} & \multicolumn{1}{|l}{4.74} \\ 
\hline
$t\overline{t}$ & 2.283 & \multicolumn{1}{|l}{3.437} & \multicolumn{1}{|l||}{
3.437} & 4.53 & \multicolumn{1}{|l}{4.75} & \multicolumn{1}{|l}{4.75} \\ 
\hline
$d\overline{d}$ & 3.119 & \multicolumn{1}{|l}{3.119} & \multicolumn{1}{|l||}{
2.054} & 4.63 & \multicolumn{1}{|l}{4.63} & \multicolumn{1}{|l}{4.37} \\ 
\hline
$s\overline{s}$ & 3.119 & \multicolumn{1}{|l}{2.054} & \multicolumn{1}{|l||}{
3.119} & 4.63 & \multicolumn{1}{|l}{4.37} & \multicolumn{1}{|l}{4.63} \\ 
\hline
$b\overline{b}$ & 2.054 & \multicolumn{1}{|l}{3.119} & \multicolumn{1}{|l||}{
3.119} & 4.37 & \multicolumn{1}{|l}{4.63} & \multicolumn{1}{|l}{4.63} \\ 
\hline\hline
$\ell ^{+}\ell ^{-}$ &  & 1.732 &  &  & 4.97 &  \\ \hline
$\nu \overline{\nu }$ &  & 1.391 &  &  & 4.80 &  \\ \hline
\end{tabular}%
\end{center}
\caption{\textit{Partial width of Z}$_{2}$ \textit{into fermions for each
representation A,B, and C in the scenery with m}$_{J_{j}}=4$ TeV\textit{.}}
\label{tab:run-quark-tree2}
\end{table}

In regard to the FCNC contributions, we found that the $Z_{1}$ flavor
changing decays are suppressed by the quadratic value of the mixing angle $%
S_{\theta }^{2}\approx 10^{-7}.$ The decays present an hierarchical order
due to the texture structure from Eqs. (\ref{texture}) and (\ref{rot-matrix}%
), such as seen in table \ref{tab:FCNC-Z-decay}. In fact, the family
structure exhibited by representation $C$ suppress most of the flavor
changing processes. We may do an estimative about the branching ratios from
table \ref{tab:FCNC-Z-decay}. For example, we get a maximum value $%
Br(Z_{1}\rightarrow s\overline{b})\approx 3\times 10^{-8}$, which is similar
to the prediction from the SM \cite{SM-FCNC} at one loop level.

Similar results are obtained for the $Z_{2}$ flavor changing decays in table %
\ref{tab:FCNC-Zprima-decay}, where the values are about one order of
magnitude bigger than the fractions obtained for the $Z_{1}$ decays. We also
evaluated the FCNC of $Z_{2}$ decay in the scenery in which $%
m_{J_{j}}\approx 4000$ GeV, obtaining the results given in table \ref%
{tab:FCNC-Zprima-decay2}. We see that the widths are slightly larger (about $%
3\%$) than the values given in table \ref{tab:FCNC-Zprima-decay} for $%
m_{J_{j}}\approx 2000$ GeV.

\begin{table}[tbp]
\begin{center}
\begin{tabular}{c|c|c}
\hline
$\Gamma _{qq^{\prime }}$ (MeV) & A-B & C \\ \hline\hline
$Z_{2}\rightarrow u\overline{c}$ & 0.0305 & 0 \\ \hline
$Z_{2}\rightarrow u\overline{t}$ & 0.0305 & 0.122 \\ \hline
$Z_{2}\rightarrow c\overline{t}$ & 2369.60 & 0 \\ \hline
$Z_{2}\rightarrow d\overline{s}$ & 107.84 & 431.35 \\ \hline
$Z_{2}\rightarrow d\overline{b}$ & 113.25 & 0 \\ \hline
$Z_{2}\rightarrow s\overline{b}$ & 2255.37 & 0 \\ \hline
\end{tabular}%
\end{center}
\caption{\textit{Partial width of Z}$_{2}$ \textit{into quarks with FCNC for
each representation A,B, and C in the scenery of m}$_{J_{j}}=4$ \textit{TeV.}
}
\label{tab:FCNC-Zprima-decay2}
\end{table}

We acknowledge the financial support from COLCIENCIAS and HELEN. R. Martinez thanks
C.P. Yuan for his hospitality in Michigan State University where part of
this work was done.

\section*{Appendix}

\appendix

\section{The Z$_{1}$-pole parameters\label{appendixAA}}

The Z$_{1}$-pole parameters with their experimental values from CERN
collider (LEP), SLAC Liner Collider (SLC) and data from atomic parity
violation taken from ref. \cite{one}, are shown in table \ref%
{tab:observables}, with the SM predictions and the expressions predicted by
331 models. The corresponding correlation matrix from ref. \cite{LEP} is
given in table \ref{tab:correlation}. For the quark masses, at Z$_{1}$-pole,
we use the following values \cite{quarks-mass}

\begin{eqnarray}
m_{u}(M_{Z_{1}}) &=&2.33_{-0.45}^{+0.42}\;\;MeV;\qquad
m_{c}(M_{Z_{1}})=677_{-61}^{+56}\;\;MeV,  \notag \\
m_{t}(M_{Z_{1}}) &=&181\pm 13\;\;GeV;\qquad
m_{d}(M_{Z_{1}})=4.69_{-0.66}^{+0.60}\;\;MeV,  \notag \\
m_{s}(M_{Z_{1}}) &=&93.4_{-13.0}^{+11.8}\;\;MeV;\qquad
m_{b}(M_{Z_{1}})=3.00\pm 0.11\;\;GeV.  \label{quarks-mass}
\end{eqnarray}

For the partial SM partial decay given by Eq. (\ref{partial-decay}), we use
the following values taken from ref. \cite{one}

\begin{eqnarray}
\Gamma _{u}^{SM} &=&0.3004\pm 0.0002\text{ }GeV;\quad \Gamma
_{d}^{SM}=0.3832\pm 0.0002\text{ }GeV;  \notag \\
\Gamma _{b}^{SM} &=&0.3758\pm 0.0001\text{ }GeV;\quad \Gamma _{\nu
}^{SM}=0.16729\pm 0.00007\text{ }GeV;  \notag \\
\Gamma _{e}^{SM} &=&0.08403\pm 0.00004\text{ }GeV.  \label{SM-partial-decay}
\end{eqnarray}

\begin{table}[tbp]
\begin{center}
$%
\begin{tabular}{c|c|c|c}
\hline
Quantity & Experimental Values & Standard Model & 331 Model \\ \hline\hline
$\Gamma _{Z}$ $\left[ GeV\right] $ & 2.4952 $\pm $ 0.0023 & 2.4972 $\pm $
0.0012 & $\Gamma _{Z}^{SM}\left( 1+\delta _{Z}\right) $ \\ \hline
$\Gamma _{had}$ $\left[ GeV\right] $ & 1.7444 $\pm $ 0.0020 & 1.7435 $\pm $
0.0011 & $\Gamma _{had}^{SM}\left( 1+\delta _{had}\right) $ \\ \hline
$\Gamma _{\left( \ell ^{+}\ell ^{-}\right) }$ $MeV$ & 83.984 $\pm $ 0.086 & 
84.024 $\pm $ 0.025 & $\Gamma _{\left( \ell ^{+}\ell ^{-}\right)
}^{SM}\left( 1+\delta _{\ell }\right) $ \\ \hline
$\sigma _{had}$ $\left[ nb\right] $ & 41.541 $\pm $ 0.037 & 41.472 $\pm $
0.009 & $\sigma _{had}^{SM}\left( 1+\delta _{\sigma }\right) $ \\ \hline
$R_{e}$ & 20.804 $\pm $ 0.050 & 20.750 $\pm $ 0.012 & $R_{e}^{SM}\left(
1+\delta _{had}+\delta _{e}\right) $ \\ \hline
$R_{\mu }$ & 20.785 $\pm $ 0.033 & 20.751 $\pm $ 0.012 & $R_{\mu
}^{SM}\left( 1+\delta _{had}+\delta _{\mu }\right) $ \\ \hline
$R_{\tau }$ & 20.764 $\pm $ 0.045 & 20.790 $\pm $ 0.018 & $R_{\tau
}^{SM}\left( 1+\delta _{had}+\delta _{\tau }\right) $ \\ \hline
$R_{b}$ & 0.21638 $\pm $ 0.00066 & 0.21564 $\pm $ 0.00014 & $%
R_{b}^{SM}\left( 1+\delta _{b}-\delta _{had}\right) $ \\ \hline
$R_{c}$ & 0.1720 $\pm $ 0.0030 & 0.17233 $\pm $ 0.00005 & $R_{c}^{SM}\left(
1+\delta _{c}-\delta _{had}\right) $ \\ \hline
$A_{e}$ & 0.15138 $\pm $ 0.00216 & 0.1472 $\pm $ 0.0011 & $A_{e}^{SM}\left(
1+\delta A_{e}\right) $ \\ \hline
$A_{\mu }$ & 0.142 $\pm $ 0.015 & 0.1472 $\pm $ 0.0011 & $A_{\mu
}^{SM}\left( 1+\delta A_{\mu }\right) $ \\ \hline
$A_{\tau }$ & 0.136 $\pm $ 0.015 & 0.1472 $\pm $ 0.0011 & $A_{\tau
}^{SM}\left( 1+\delta A_{\tau }\right) $ \\ \hline
$A_{b}$ & 0.925 $\pm $ 0.020 & 0.9347 $\pm $ 0.0001 & $A_{b}^{SM}\left(
1+\delta A_{b}\right) $ \\ \hline
$A_{c}$ & 0.670 $\pm $ 0.026 & 0.6678 $\pm $ 0.0005 & $A_{c}^{SM}\left(
1+\delta A_{c}\right) $ \\ \hline
$A_{s}$ & 0.895 $\pm $ 0.091 & 0.9357 $\pm $ 0.0001 & $A_{s}^{SM}\left(
1+\delta A_{s}\right) $ \\ \hline
$A_{FB}^{\left( 0,e\right) }$ & 0.0145 $\pm $ 0.0025 & 0.01626 $\pm $ 0.00025
& $A_{FB}^{(0,e)SM}\left( 1+2\delta A_{e}\right) $ \\ \hline
$A_{FB}^{\left( 0,\mu \right) }$ & 0.0169 $\pm $ 0.0013 & 0.01626 $\pm $
0.00025 & $A_{FB}^{(0,\mu )SM}\left( 1+\delta A_{e}+\delta A_{\mu }\right) $
\\ \hline
$A_{FB}^{\left( 0,\tau \right) }$ & 0.0188 $\pm $ 0.0017 & 0.01626 $\pm $
0.00025 & $A_{FB}^{(0,\tau )SM}\left( 1+\delta A_{e}+\delta A_{\tau }\right) 
$ \\ \hline
$A_{FB}^{\left( 0,b\right) }$ & 0.0997 $\pm $ 0.0016 & 0.1032 $\pm $ 0.0008
& $A_{FB}^{(0,b)SM}\left( 1+\delta A_{e}+\delta A_{b}\right) $ \\ \hline
$A_{FB}^{\left( 0,c\right) }$ & 0.0706 $\pm $ 0.0035 & 0.0738 $\pm $ 0.0006
& $A_{FB}^{(0,c)SM}\left( 1+\delta A_{e}+\delta A_{c}\right) $ \\ \hline
$A_{FB}^{\left( 0,s\right) }$ & 0.0976 $\pm $ 0.0114 & 0.1033 $\pm $ 0.0008
& $A_{FB}^{(0,s)SM}\left( 1+\delta A_{e}+\delta A_{s}\right) $ \\ \hline
$Q_{W}(Cs)$ & $-$72.69 $\pm $ 0.48 & $-$73.19 $\pm $ 0.03 & $%
Q_{W}^{SM}\left( 1+\delta Q_{W}\right) $ \\ \hline
\end{tabular}%
\ \ $%
\end{center}
\caption{\textit{The parameters for experimental values, SM predictions and
331 corrections. The values are taken from ref. \protect\cite{one}}}
\label{tab:observables}
\end{table}

\begin{table}[tbp]
\begin{tabular}{ll}
\hline
$\Gamma _{had}$ & $\Gamma _{\ell }$ \\ \hline\hline
1 &  \\ 
.39 & 1 \\ \hline
\end{tabular}%
\par
\begin{tabular}{lll}
\hline
$A_{e}$ & $A_{\mu }$ & $A_{\tau }$ \\ \hline\hline
1 &  &  \\ 
.038 & 1 &  \\ 
.033 & .007 & 1 \\ \hline
\end{tabular}%
\par
\begin{tabular}{llllll}
\hline
$R_{b}$ & $R_{c}$ & $A_{b}$ & $A_{c}$ & $A_{FB}^{(0,b)}$ & $A_{FB}^{(0,c)}$
\\ \hline\hline
1 &  &  &  &  &  \\ 
-.18 & 1 &  &  &  &  \\ 
-.08 & .04 & 1 &  &  &  \\ 
.04 & -.06 & .11 & 1 &  &  \\ 
-.10 & .04 & .06 & .01 & 1 &  \\ 
.07 & -.06 & -.02 & .04 & .15 & 1 \\ \hline
\end{tabular}%
\par
\begin{tabular}{llllllll}
\hline
$\Gamma _{Z}$ & $\sigma _{had}$ & $R_{e}$ & $R_{\mu }$ & $R_{\tau }$ & $%
A_{FB}^{(0,e)}$ & $A_{FB}^{(0,\mu )}$ & $A_{FB}^{(0,\tau )}$ \\ \hline\hline
1 &  &  &  &  &  &  &  \\ 
-.297 & 1 &  &  &  &  &  &  \\ 
-.011 & .105 & 1 &  &  &  &  &  \\ 
.008 & .131 & .069 & 1 &  &  &  &  \\ 
.006 & .092 & .046 & .069 & 1 &  &  &  \\ 
.007 & .001 & -.371 & .001 & .003 & 1 &  &  \\ 
.002 & .003 & .020 & .012 & .001 & -.024 & 1 &  \\ 
.001 & .002 & .013 & -.003 & .009 & -.020 & .046 & 1 \\ \hline
\end{tabular}%
\caption{\textit{The correlation coefficients for the Z-pole observables}}
\label{tab:correlation}
\end{table}

\section{Radiative corrections\label{appendixA}}

The $Z_{1}$ and $Z_{2}$ decay in Eqs. (\ref{partial-decay}) and (\ref%
{Z'-width}) contains global QED and QCD corrections through the definition
of $R_{QED}=1+\delta _{QED}^{f}$ and $R_{QCD}=1+\frac{1}{2}\left(
N_{c}^{f}-1\right) \delta _{QCD}^{f},$ where \cite{one, pitch}

\begin{eqnarray}
\delta _{QED}^{f} &=&\frac{3\alpha Q_{f}^{2}}{4\pi };  \notag \\
\delta _{QCD}^{f} &=&\frac{\alpha _{s}}{\pi }+1.405\left( \frac{\alpha _{s}}{%
\pi }\right) ^{2}-12.8\left( \frac{\alpha _{s}}{\pi }\right) ^{3}-\frac{%
\alpha \alpha _{s}Q_{f}^{2}}{4\pi ^{2}}  \label{QCD}
\end{eqnarray}%
with $\alpha $ and $\alpha _{s}$ the electromagnetic and QCD constants,
respectively. The values $\alpha $ and $\alpha _{s}$ are calculated at the $%
M_{Z_{1}}$ scale for the $Z_{1}$ decays, and at the $M_{Z_{2}}$ scale for
the $Z_{2}$ decays.

We are also considering oblique corrections sensitive to the top quark mass
in the $Z_{1}$ decay. Here, we show the calculation of the oblique
corrections corresponding to the $Z_{2}$ decay, which is mostly sensitive to
the extra quarks masses $m_{J_{1,2,3}}.$ The correction due to the $Z_{2}$
self-energy leads to the wavefunction renormalization%
\begin{equation}
Z_{2}\rightarrow Z_{2R}\approx \left( 1-\frac{1}{2}\Sigma
_{Z_{2}Z_{2}}^{(fin)\prime }(q^{2})\right) Z_{2},  \label{Z'-renormalizado}
\end{equation}

\noindent where the finite part of the self-energy gives

\begin{eqnarray}
\Sigma _{Z_{2}Z_{2}}^{(fin)}(q^{2}) &\approx &\frac{1}{12\pi ^{2}}\left( 
\frac{g}{2C_{W}}\right) ^{2}\sum_{j=1}^{3}\left\{ \left( \widetilde{g}%
_{v}^{J_{j}}\right) ^{2}\left[ -q^{2}\ln \frac{m_{J_{j}}^{2}}{q^{2}}-\frac{%
q^{2}}{3}\right] \right.  \notag \\
&&+\left. \left( \widetilde{g}_{a}^{J_{j}}\right) ^{2}\left[ \left(
6m_{J_{j}}^{2}-q^{2}\right) \ln \frac{m_{J_{j}}^{2}}{q^{2}}-\frac{q^{2}}{3}%
\right] \right\} ,  \label{Z'-Z'-self}
\end{eqnarray}

\noindent and

\begin{eqnarray}
\Sigma _{Z_{2}Z_{2}}^{(fin)^{\prime }}(q^{2}) &=&\frac{d\Sigma
_{Z_{2}Z_{2}}^{(fin)}}{dq^{2}}=\frac{1}{12\pi ^{2}}\left( \frac{g}{2C_{W}}%
\right) ^{2}\sum_{j=1}^{3}\left\{ \left( \widetilde{g}_{v}^{J_{j}}\right)
^{2}\left( \frac{2}{3}-\ln \frac{m_{J_{j}}^{2}}{q^{2}}\right) \right.  \notag
\\
&&+\left. \left( \widetilde{g}_{a}^{J_{j}}\right) ^{2}\left( \frac{2}{3}-\ln 
\frac{m_{J_{j}}^{2}}{q^{2}}-\frac{6m_{J_{j}}^{2}}{q^{2}}\right) \right\} .
\label{derivative-Z'-Z'-self}
\end{eqnarray}

\noindent The $Z_{2}-Z_{1}$ self-energy leads to the following vacuum
polarization

\begin{eqnarray}
\Pi _{Z_{2}Z_{1}}^{(fin)}(q^{2}) &\approx &\frac{1}{12\pi ^{2}}\left( \frac{g%
}{2C_{W}}\right) ^{2}\sum_{j=1}^{3}\left\{ \widetilde{g}%
_{v}^{J_{j}}g_{v}^{J_{j}}\left[ -\ln \frac{m_{J_{j}}^{2}}{q^{2}}-\frac{1}{3}%
\right] \right.  \notag \\
&&+\left. \widetilde{g}_{a}^{J_{j}}g_{a}^{J_{j}}\left[ \left( \frac{%
6m_{J_{j}}^{2}}{q^{2}}-1\right) \ln \frac{m_{J_{j}}^{2}}{q^{2}}-\frac{1}{3}%
\right] \right\} ,  \label{Z'-Z-self}
\end{eqnarray}

\noindent and the $Z_{2}$-photon vacuum polarization is given by

\begin{equation}
\Pi _{Z_{2}\gamma }^{(fin)}(q^{2})\approx \frac{1}{12\pi ^{2}}\frac{%
g^{2}S_{W}}{2C_{W}}\sum_{j=1}^{3}\left\{ Q_{J_{j}}\widetilde{g}_{v}^{J_{j}}%
\left[ -\ln \frac{m_{J_{j}}^{2}}{q^{2}}-\frac{1}{3}\right] \right\} ,
\label{Z'-photon-self}
\end{equation}

\noindent with $Q_{J_{j}}$ the electric charge of the virtual $J_{j}$ quarks
given in table \ref{tab:espectro}.

\section{Running masses and coupling constants\label{appendixB}}

The solution of the renormalization group at the lowest one-loop order gives
the running coupling constant for $\widetilde{M}\leq \mu $%
\begin{equation}
g_{i}^{-2}(\widetilde{M})=g_{i}^{-2}(\mu )+\frac{b_{i}}{8\pi ^{2}}\ln \left( 
\frac{\mu }{\widetilde{M}}\right) ,  \label{runn-coup-const}
\end{equation}

\noindent for $i=1,2,3,$ each one corresponding to the constant coupling of $%
U(1)_{Y},$ $SU(2)_{L}$ and $SU(3)_{c},$ respectively. Specifically, we use
the matching condition for the constant couplings, where the $SU(3)_{L}$
constant is the same as the $SU(2)_{L}$ constant$,$ i.e. $g_{2}=g.$ Running
the constants at the scale $\mu =M_{Z_{2}}$ and taking $\widetilde{M}%
=M_{Z_{1}},$ we obtain for $g_{1}$ and $g_{2}$

\begin{eqnarray}
g_{Y}^{2}(M_{Z_{2}}) &=&\frac{g_{Y}^{2}(M_{Z_{1}})}{1-\frac{b_{1}}{8\pi ^{2}}%
g_{Y}^{2}(M_{Z_{1}})\ln \left( \frac{M_{Z_{2}}}{M_{Z_{1}}}\right) }  \notag
\\
g^{2}(M_{Z_{2}}) &=&\frac{g^{2}(M_{Z_{1}})}{1-\frac{b_{2}}{8\pi ^{2}}%
g^{2}(M_{Z_{1}})\ln \left( \frac{M_{Z_{2}}}{M_{Z_{1}}}\right) }
\label{g-running}
\end{eqnarray}

\noindent with

\begin{eqnarray}
b_{1} &=&\frac{20}{9}N_{g}+\frac{1}{6}N_{H}+\frac{1}{3}\sum_{\text{sing}%
}Y^{2}=\frac{22}{3},  \notag \\
b_{2} &=&\frac{4}{3}N_{g}+\frac{1}{6}N_{H}+\frac{22}{3}=-3,
\label{renorm-coef}
\end{eqnarray}

\noindent where $N_{g}=3$ is the number of fermion families and $N_{H}=2$
the number of $SU(2)_{L}$ scalar doublets. With the above definitions, we
can obtain the running Weinberg angle

\begin{eqnarray}
S_{W}^{2}(M_{Z_{2}}) &=&\frac{g_{Y}^{2}(M_{Z_{2}})}{%
g^{2}(M_{Z_{2}})+g_{Y}^{2}(M_{Z_{2}})}  \notag \\
&=&S_{W}^{2}(M_{Z_{1}})\left[ \frac{1-\frac{b_{2}}{2\pi }\alpha
_{2}(M_{Z_{1}})\ln \left( M_{Z_{2}}/M_{Z_{1}}\right) }{1-\frac{b_{1}+b_{2}}{%
2\pi }\alpha (M_{Z_{1}})\ln \left( M_{Z_{2}}/M_{Z_{1}}\right) }\right] .
\label{sw-running}
\end{eqnarray}

In order to calculate the running masses for all quarks, we should use the
running QCD constant at the $n$th quark threshold \cite{quarks-mass}, which
is defined as

\begin{eqnarray}
\alpha _{s}^{(n)}(\mu ) &=&\frac{4\pi }{\beta _{0}^{(n)}L^{(n)}}\left\{ 1-%
\frac{2\beta _{1}^{(n)}\ln \left[ L^{(n)}\right] }{\left( \beta
_{0}^{(n)}\right) ^{2}L^{(n)}}+\frac{4\left( \beta _{1}^{(n)}\right) ^{2}}{%
\left( \beta _{0}^{(n)}\right) ^{4}\left( L^{(n)}\right) ^{2}}\right.  \notag
\\
&&\times \left. \left[ \left( \ln \left( L^{(n)}\right) -\frac{1}{2}\right)
^{2}+\frac{\beta _{0}^{(n)}\beta _{2}^{(n)}}{8\left( \beta _{1}^{(n)}\right)
^{2}}-\frac{5}{4}\right] \right\} ,  \label{strong-running}
\end{eqnarray}%
with $L^{(n)}=\ln \left( \frac{\mu ^{2}}{\left( \Lambda ^{(n)}\right) ^{2}}%
\right) ,$ $\beta _{0}^{(n)}=11-\frac{2}{3}n_{f},$ $\beta _{1}^{(n)}=51-%
\frac{19}{3}n_{f},$ $\beta _{2}^{(n)}=2857-\frac{5033}{9}n_{f}+\frac{325}{27}%
n_{f}^{2},$ and $n_{f}$ the number of quarks with mass less than $\mu .$ The
asymptotic scale parameters $\Lambda ^{(n)}$ for the energy scale $\mu $ at
each quark threshold are determined by \cite{lambda}

\begin{eqnarray}
2\beta _{0}^{(n-1)}\ln \left( \frac{\Lambda ^{(n)}}{\Lambda ^{(n-1)}}\right)
&=&\left( \beta _{0}^{(n)}-\beta _{0}^{(n-1)}\right) L^{(n)}+2\left( \frac{%
\beta _{1}^{(n)}}{\beta _{0}^{(n)}}-\frac{\beta _{1}^{(n-1)}}{\beta
_{0}^{(n-1)}}\right) \ln \left( L^{(n)}\right)  \notag \\
&&-\frac{2\beta _{1}^{(n-1)}}{\beta _{0}^{(n-1)}}\ln \left( \frac{\beta
_{0}^{(n)}}{\beta _{0}^{(n-1)}}\right) +\frac{4\beta _{1}^{(n)}}{\left(
\beta _{0}^{(n)}\right) ^{2}}\left( \frac{\beta _{1}^{(n)}}{\beta _{0}^{(n)}}%
-\frac{\beta _{1}^{(n-1)}}{\beta _{0}^{(n-1)}}\right) \frac{\ln \left(
L^{(n)}\right) }{L^{(n)}}  \notag \\
&&+\frac{1}{\beta _{0}^{(n)}}\left[ \left( \frac{2\beta _{1}^{(n)}}{\beta
_{0}^{(n)}}\right) ^{2}-\left( \frac{2\beta _{1}^{(n-1)}}{\beta _{0}^{(n-1)}}%
\right) ^{2}-\frac{\beta _{2}^{(n)}}{2\beta _{0}^{(n)}}+\frac{\beta
_{2}^{(n-1)}}{2\beta _{0}^{(n-1)}}-\frac{22}{9}\right] \frac{1}{L^{(n)}}, 
\notag \\
&&  \label{Lambda}
\end{eqnarray}%
where the starting parameter is $\Lambda ^{(5)}=217_{-23}^{+25}$ MeV \cite%
{one}. We get for each threshold $\mu =m_{q}^{(n)}$ (with $n=3$ for the
light quarks $u,d,s$ below 1 GeV; and $n=4,5,6$, each one corresponding to
the heavy quarks $c,b$ and $t,$ respectively)

\begin{equation}
\Lambda ^{(3)}=342\text{ MeV};\qquad \Lambda ^{(4)}=301\text{ MeV};\qquad
\Lambda ^{(6)}=91.7\text{ MeV.}\qquad
\end{equation}

The running mass for the heavy quarks $q=c,b,t$ at $\mu <\mu ^{n+1}$ is 
\begin{equation}
\overline{m}_{qn}(\mu )=\frac{R^{(n)}(\mu )}{R^{(n)}(m^{pole})}m_{q}^{pole},
\label{run-heavy-mass}
\end{equation}%
\ while for the light quarks $q=u,d,s$ is

\begin{equation}
\overline{m}_{q}(\mu )=\frac{R^{(3)}(\mu )}{R^{(3)}(1\text{ GeV})}m_{q}(1%
\text{ GeV}),  \label{run-light-mass}
\end{equation}%
where $m_{q}^{pole}$ are the pole masses and $m_{q}(1$ GeV$)$ are the masses
measured at $1$ GeV scale. We use the following values \cite{quarks-mass}

\begin{eqnarray}
m_{c}^{pole} &=&1.26\pm 0.13\;GeV,\qquad m_{b}^{pole}=4.26\pm 0.15\;GeV, 
\notag \\
m_{t}^{pole} &=&174.3\pm 5.1\;GeV,\qquad m_{u}(1\text{ GeV})=4.88\pm
0.57\;MeV,  \notag \\
m_{d}(1\text{ GeV}) &=&9.81\pm 0.65\;MeV,\qquad m_{s}(1\text{ GeV})=195.4\pm
12.5\;MeV.  \label{pole-mass}
\end{eqnarray}

We also use

\begin{eqnarray}
R^{(n)}(\mu ) &=&\left( \frac{\beta _{0}\alpha _{s}^{(n)}}{2\pi }\right)
^{2\gamma _{0}/\beta _{0}}\left\{ 1+\left[ \frac{2\gamma _{1}}{\beta _{0}}-%
\frac{\beta _{1}\gamma _{0}}{\beta _{0}^{2}}\right] \frac{\alpha _{s}^{(n)}}{%
\pi }\right.  \notag \\
&&+\left. \frac{1}{2}\left[ \left( \frac{2\gamma _{1}}{\beta _{0}}-\frac{%
\beta _{1}\gamma _{0}}{\beta _{0}^{2}}\right) ^{2}+\frac{2\gamma _{2}}{\beta
_{0}}-\frac{\beta _{1}\gamma _{1}}{\beta _{0}^{2}}-\frac{\beta _{2}\gamma
_{0}}{16\beta _{0}^{2}}+\frac{\beta _{1}^{2}\gamma _{0}}{2\beta _{0}^{3}}%
\right] \left( \frac{\alpha _{s}^{(n)}}{\pi }\right) ^{2}\right\} ,
\label{R}
\end{eqnarray}%
with $\gamma _{0}=2,$ $\gamma _{1}=\frac{101}{12}-\frac{5}{18}n_{f},$ and $%
\gamma _{2}=\frac{1}{32}\left[ 1249-\left( \frac{2216}{27}+\frac{160}{3}%
\zeta (3)\right) n_{f}-\frac{140}{81}n_{f}^{2}\right] .$ In order to get the
running mass at $\mu =M_{Z_{2}}>\mu ^{(6)}=m_{t}$, it is necessary to use
the matching condition \cite{lambda} 
\begin{equation}
\overline{m}_{qn}^{(N)}(\mu )=\overline{m}_{qn}^{(N-1)}(\mu )\left[ 1+\frac{1%
}{12}\left( x_{N}^{2}+\frac{5}{3}x_{N}+\frac{89}{36}\right) \left( \frac{%
\alpha _{s}^{(N)}}{\pi }\right) ^{2}\right] ^{-1},  \label{matching}
\end{equation}%
where $x_{N}=\ln \left[ \left( m_{q_{N}}^{(N)}/\mu \right) ^{2}\right] $, $%
N>n,$ and $\mu _{N}\leq \mu \leq \mu _{N+1}.$ By iterating the above
equation, along with the definitions (\ref{run-heavy-mass}) and (\ref%
{run-light-mass}) at each quark threshold, we obtain the values given by Eq.
(\ref{running-quarks-mass}).

\newpage

\end{document}